# Origin of nucleosynthetic isotope variability in the NC reservoir: Evidence from Ti, Cr, and Mo isotopes

**Elias Wölfer[a, b, *], Christoph Burkhardt[a, b], Gerrit Budde[b], Christian A. Jansen[b], Jonas Pape[b], and Thorsten Kleine[a]**

[a]Max Planck Institute for Solar System Research, Justus-von-Liebig-Weg 3, 37077 Göttingen, Germany.

[b] Institut für Planetologie, University of Münster, Wilhelm-Klemm-Str. 10, 48149 Münster, Germany.

[*]corresponding author: woelfer@mps.mpg.de

**Accepted for publication in**
***Geochimica et Cosmochimica Acta***

**Abstract**

Nucleosynthetic isotope anomalies allow distinguishing between non-carbonaceous (NC) and carbonaceous (CC) type meteorites, and have revealed correlated isotope variations especially among NC bodies. Understanding the origin of this NC trend is important for identifying the processes that produced the NC isotope heterogeneity, and for using these isotope anomalies to reconstruct the early evolution of the solar protoplanetary disk. We report mass-independent Ti, Cr, and Mo isotope compositions for a comprehensive set of previously not or only poorly investigated meteorites, as well as acid leachates obtained from the sequential digestion of primitive ordinary chondrites. Some of the samples investigated in this study fill previously identified apparent gaps in the NC trend, suggesting these gaps reflect unrepresentative sampling of a more continuous isotopic trend. Bulk meteorites and leachates exhibit distinct isotope systematics, indicating that the NC isotope variability does not reflect selective thermal processing of presolar carriers in the disk. The NC trend also cannot reflect the continuous addition of CC dust from the outer to the inner disk, because early- and late-formed NC meteorites display largely overlapping isotopic compositions. Instead, we find that the NC isotope heterogeneity is best accounted for by fractionation and mixing among chemically and isotopically distinct dust components, similar to the processes that produced the isotopic variability among carbonaceous chondrites. On this basis we argue for the presence of substructures in the inner disk, which facilitated fractionation and mixing among distinct dust components, and helped preserve a long-lived dust reservoir from which NC planetesimals accreted over an extended period of time.

## 1 Introduction

Nucleosynthetic isotope anomalies allow distinguishing between non-carbonaceous (NC) and carbonaceous chondrite (CC) type meteorites, which have been suggested to derive from bodies formed in the inner and outer solar accretion disk, respectively. These anomalies result from the heterogeneous distribution of isotopically anomalous matter derived from different nucleosynthetic sources, and occur among individual components of primitive meteorites as well as at the bulk meteorite and planetary scale. This makes these isotope anomalies powerful tracers of genetic relationships among and between meteorites and their components, of the extent of mixing and fractionation of dust components in the solar nebula prior to parent body accretion, and of the provenance of planetary building materials (see reviews by Kleine et al., 2020; Kruijer et al., 2020; Bermingham et al., 2020; Kleine and Nimmo, 2025; Tissot et al., 2025).

Nucleosynthetic isotope anomalies in Ti, Cr, and Mo are particularly important because they have been instrumental for defining the NC-CC dichotomy (e.g., Trinquier et al., 2007; Warren, 2011a; Budde et al., 2016). A large body of isotopic data exists for these three elements, including data for bulk meteorites as well components of primitive chondrites (e.g., Trinquier et al., 2007, 2009; Qin et al., 2010; Zhang et al., 2012; Budde et al., 2016, 2019; Gerber et al., 2017; Spitzer et al., 2020; Williams et al., 2020; Schneider et al., 2020; Zhu et al., 2021b, 2023; Rüfenacht et al., 2023; Yokoyama et al., 2023; Jansen et al., 2024). These data, in addition to defining the NC-CC dichotomy, reveal correlated isotopic variations within each reservoir (e.g., Yokoyama et al., 2019; Spitzer et al., 2020; Burkhardt et al., 2021; Render et al., 2022; Zhu et al., 2023; Tissot et al., 2025). Among the carbonaceous chondrites, these correlated isotope variations are thought to reflect varying abundances of their constituent components, namely refractory inclusions, chondrules, matrix, and Fe-Ni metal (Alexander, 2019; Hellmann et al., 2023; Spitzer et al., 2025b; Gurrutxaga et al., 2026). However, the origin of isotopic variations among the NC meteorites is less well understood. They have been suggested to reflect (*i*) thermal processing of presolar components (Trinquier et al., 2009; Ek et al., 2020), (*ii*) a temporal evolution of the inner disk's isotopic composition through the continuous addition of inward-drifting CI chondrite-like pebbles (e.g., Schiller et al., 2018), and (*iii*) mixing between two primordial disk reservoirs (e.g., Spitzer et al., 2020; Burkhardt et al., 2021).

Although Ti, Cr, and Mo isotopic data exist for many different meteorite groups, several groups especially from the NC reservoir, and a large number of ungrouped meteorites, have not or only poorly been investigated, and lack data for one or several of these elements. To better understand the full range of isotopic variations within the NC reservoir and, ultimately, the origin of isotope variability within the inner protoplanetary disk, we obtained Ti, Cr, and Mo isotopic data for a comprehensive set of previously not, or only scarcely, investigated NC and some CC meteorites. In addition, we present the first Ti isotopic data for acid leachates obtained by the sequential digestion of ordinary chondrites, for which Mo isotopic data have been reported previously (Budde et al., 2019). The leachate data provide information on the internal isotopic heterogeneity among the components of primitive chondrites and, as such, on the link between this internal isotopic variability and the isotope variations among bulk meteorites (e.g., Trinquier et al., 2009; Burkhardt et al., 2019). Altogether, the new data provide new constraints on genetic relationships among NC bodies and the origin and nature of the correlated isotope variations among inner solar system objects.

## 2 Materials and methods

### *2.1 Samples*

A total of 66 samples have been selected for this study, of which 38 have an NC and 28 a CC heritage (Table 1). We report Ti isotopic data for all of these samples, Cr isotope data for 22 of these samples, and Mo isotopic data for five of these samples. Further Mo isotope data for 26 of these samples have been reported in Worsham et al. (2017), Budde et al. (2019), and Hopp et al. (2020). All measurements have been performed on aliquots of the same sample powders or digestion solutions (Table 1).

The bulk NC meteorite samples include various achondrites (i.e. acapulcoite-lodranites, brachinites, mesosiderites, ureilites, aubrites, angrites, HEDs, winonaites), Rumuruti chondrites (RC), and one Kakangari chondrite (KC) (Table 1). In addition, two ungrouped chondrites (NWA 5492, NWA 13202) and six ungrouped NC achondrites (NWA 1058, NWA 2526, NWA 5400, NWA 6112, GRA 06128) were analyzed. The new data substantially expand the existing Ti-Cr-Mo isotope data set for NC meteorites, including samples for which the genetic association to specific groups of meteorites is uncertain. For instance, we report the first Ti and Mo and only the second Cr isotopic data for KCs (Zhu et al., 2023). We also report the first Mo isotopic data for ungrouped NC chondrites (NWA 13202, NWA 5492) and the ungrouped NC achondrite (GRA 06128). Three samples (GRA 06128, NWA 5400, NWA 6112) feature chemical and petrologic similarities to brachinites (Day et al., 2012; Hasegawa et al., 2019), but their genetic association to this meteorite group is unknown. The unique primitive achondrite NWA 1058 shares chemical similarities with acapulcoite-lodranites and has also been classified as a winonaite or metal-rich diogenite, but its O isotope composition and cosmic ray exposure history is different from these meteorite groups (Eugster and Lorenzetti, 2005). The unique ungrouped metal-rich chondrite NWA 13202 is paired with NWA 12273 and NWA 12379 and shows petrographic similarities (e.g., lack of matrix, 70 vol.% metal) to metal-rich CB chondrites and G chondrite grouplet (Jansen et al., 2019). However, multiple other characteristics of NWA 13202, including chondrule sizes and textures, as well as O isotopic and chemical compositions of chondrules, are more similar to ordinary chondrites. Finally, the ungrouped metal-rich chondrite NWA 5492 has been related to the metal-rich chondrites of the G chondrite grouplet (Weisberg et al., 2015; Ivanova et al., 2020), but no nucleosynthetic isotope data have been reported for samples of this grouplet, except for one Cr isotopic measurement for GRO 95551 (Zhu et al., 2023).

The bulk CC meteorite samples of this study include (*i*) five samples of the major chondrite groups (CV, CM, CL), (*ii*) twelve, petrographically heterogeneous subsamples of the CV-type Allende chondrite, (*iii*) several ungrouped chondrites, some of which have previously been investigated for Cr and Ti isotopes (Hellmann et al., 2023), (*iv*) and one ungrouped achondrite (NWA 6926) (Table 2). The twelve Allende subsamples are characterized by different proportions of chondrules, matrix, and refractory inclusions and stem from powders originally prepared by Stracke et al. (2012) from 7×7×4 mm (~620 mg) cubes cut from a large slice of Allende. These samples have previously been analyzed for their chemical compositions and we here report their Ti isotopic compositions. We report the first Mo isotopic data for recently established group of CL chondrites (Metzler et al., 2021), the first Ti isotopic data for some ungrouped carbonaceous chondrites (DaG 055, DaG 430), and the first Ti and Cr isotopic data for the ungrouped achondrite NWA 6926.

We also obtained Ti isotopic data for six leachate fractions each from two primitive ordinary chondrites that have previously been analyzed for their Mo isotopic compositions (Budde et al., 2019) (Table 3). The leachates were obtained by the sequential (six-step) digestion of sample powders of NWA 2458 (L3.2; 6.48 g) and WSG 95300 (H3.3; 4.35 g), with the strength of the acids successively increasing in the sequence (Table S1; Budde et al., 2019). As such, easily dissolvable mineral phases are enriched in the leaching steps L1 and L2, major silicate phases are dissolved in L3 and L4, whereas more acid-resistant, refractory phases are enriched in L5 and L6.

### *2.2 Sample preparation*

Bulk meteorite samples (~0.5–1 g) were powdered in an agate mortar and aliquots of each sample were digested in concentrated HF-$HNO_3$-$HClO_4$ (2:1:0.05) at 180–200 °C (5 days), followed by aqua regia (3:1 concentrated HCl-$HNO_3$) at 130–150 °C (2 days). The sample solutions of the investigated ureilites still contained finely dispersed particles (e.g., graphite, spinel, rutile, ilmenite, titanite, and carbides; Warren, 2011b; Budde et al., 2015) after table-top digestion. Thus, the residues were further digested in concentrated HF-$HNO_3$ (1:1) in steel-jacketed teflon Parr bombs at 190 °C (4 days). After this step, the residual phases (e.g., spinels) were mostly in solution, but some finely dispersed material (e.g., graphite or carbides) were left, which potentially could affect the Ti isotopic composition of the bulk ureilites in the form of Ti-carbides (van Helden et al., 2000). For this reason, the samples were further digested using inverse aqua regia inside sealed Carius tubes at 230 °C (3 days) (Shirey and Walker, 1995). However, those ureilites that have been dissolved via table-top digestion (Budde et al., 2015, 2019) show indistinguishable Ti and Cr isotopic compositions compared to those that have been digested by additional treatment in Parr bombs and Carius tubes (this study). Hence, the potential incomplete dissolution of acid-resistant material during table-top digestions of ureilites is inconsequential for their measured Ti and Cr isotope anomalies.

For samples previously analyzed for Mo by Worsham et al. (2017), Budde et al. (2019), and Hopp et al. (2020), Ti was collected during the clean-up step of the two-stage anion exchange chromatography used for the separation of W, where Ti is eluted in 1 M HCl–2 % $H_2O_2$. Aliquots (equivalent to ~30 µg Ti) were taken from these solutions and treated with *aqua regia* to destroy any organic compounds that might have formed during the W chemistry. Finally, Ti concentrations for all sample solutions were measured on small aliquots (0.5–1 vol.%) on a ThermoScientific XSeries II quadrupole ICP-MS at the University of Münster.

## *2.3 Chemical separation and isotope measurements*

The analytical procedure described below was the same for the newly digested samples as well as the aliquots taken from previous sample digestion from the studies of Worsham et al. (2017), Budde et al. (2015, 2019), and Hopp et al. (2020). Titanium was separated from the sample matrix by two-stage anion exchange chromatography, following previously established protocols (Zhang et al., 2011; Torrano et al., 2019). Yields were typically 90–95%, and procedural blanks were negligible throughout. The highest blanks were observed for the digestion of the ureilites as described above. These blanks were as high as ~10 ng Ti, which is still negligible compared to the 65–115 µg Ti processed per sample.

The Ti isotope measurements were performed on a ThermoScientific Neptune *Plus* MC-ICP-MS in the Institut für Planetologie at the University of Münster. The analytical protocol was adapted from Zhang et al. (2011), Gerber et al. (2017), and Burkhardt et al. (2019). Sample solutions of ~600 ng/g Ti in 0.3 M $HNO_3$–0.0014 M HF were introduced using a Savillex nebulizer (~60 µl/min uptake rate) and a Cetac Aridus II desolvator. With this setup, ion beam intensities of $\sim3.5\times10^{-10}$ A on $^{48}Ti$ were obtained in high resolution mode. Instrumental mass bias was corrected by internal normalization to $^{49}Ti/^{47}Ti = 0.749766$ using the exponential law. The Ti isotope composition is reported in the ε-notation (i.e. 0.01%) relative to bracketing analyses of the Origins Lab OL-Ti solution standard. For all samples, the reported data represent the mean of repeat measurements and the stated uncertainties reflect Student-t 95% confidence intervals (95% CI). The accuracy and reproducibility of the Ti isotope measurements were assessed by repeated analyses of geochemical reference basalts (BHVO-2 and JB-2), which were processed through the full chemical separation and analyzed together with each set of samples. The $\varepsilon^{i}Ti$ values obtained for the terrestrial rock standard JB-2 are indistinguishable from the OL-Ti solution standard (Millet and Dauphas, 2014), while repeated measurements of several digestions of BHVO-2 display small offsets of about –0.1 $\varepsilon^{50}Ti$, consistent with results of previous studies (Gerber et al., 2017; Render et al., 2019; Torrano et al., 2019, 2024; Rüfenacht et al., 2023).

Chromium was collected during the first step of the anion exchange chromatography used for the separation of Ti, where Cr is eluted in 12 M $HNO_3$ together with most other matrix elements. For samples that have previously been analyzed for Mo isotopes, Cr was collected using 0.5 M HCl–0.5 M HF during the first step of the anion exchange chromatography used for the separation of W and Mo (Budde et al., 2019; Hopp et al., 2020). Chromium concentrations for all samples were measured on small aliquots (0.1 vol.%) on a ThermoScientific XSeries II quadrupole ICP-MS at the University of Münster, and finally, aliquots (equivalent to ~30 µg Cr) were taken from the matrix cuts of either the Ti or the W/Mo chemistry. Chromium was separated from matrix elements using a combination of anion and cation exchange columns (Schneider et al., 2020). The Cr yields as determined by ICPMS were typically ~80% and total procedural blanks (<20 ng) were negligible. The Cr isotope measurements were performed on the ThermoScientific Triton *Plus* TIMS in the Institut für Planetologie following the method described in Schneider et al. (2020). Each measurement consisted of a four-line data acquisition scheme, where all Cr isotopes as well as Ti, V, and Fe interference monitors on masses 49, 51, and 56 were measured. Instrumental mass bias was corrected by internal normalization to $^{50}Cr/^{52}Cr = 0.051859$ and using the exponential law. Measurements were performed with a stable ion beam of $\sim 1 \times 10^{-10}$ A on $^{52}Cr$ for >4 hours. The Cr isotope composition of a sample is given in the ε-notation relative to the session average of several runs of the NIST SRM3112a solution standard. For all samples, the reported data represent the mean of repeat measurements and the stated uncertainties reflect 95% CI. The accuracy and reproducibility of the Cr isotope measurements were assessed by analyses of terrestrial reference materials (BHVO-2, JA-2, DTS-2b), which were processed through the full chemical separation described above and analyzed together with each set of samples. The $\varepsilon^{i}Cr$ values of these rocks are indistinguishable within uncertainty and display slightly elevated $\varepsilon^{54}Cr$ values of between ~0.07 and ~0.20 (Table 1), consistent with results of previous studies (Trinquier et al., 2007; Qin et al., 2010; Mougel et al., 2018; Zhu et al., 2021b; Xu et al., 2023). This small offset likely reflects a non-exponential isotope fractionation of Cr in the standard.

Molybdenum isotopes have been measured for LEW 87232 (KC), NWA 5492 (GC), GRA 06128 (ungrouped achondrite), NWA 13400 (CL), and a metal separate of NWA 13202 (ungrouped chondrite). For these samples, Mo was separated from the sample matrix via anion exchange chromatography, following our previously established procedures (Budde et al., 2019, and references therein). Yields as measured by quadrupole ICP-MS were typically ~75%, and procedural blanks were negligible. The Mo isotope measurements were performed on the ThermoScientific Neptune *Plus* MC-ICP-MS in the Institut für Planetologie. Sample solutions of ~100 ng/g Mo were introduced through a Savillex nebulizer with an ~50 µl/min uptake rate attached to a Aridus II desolvating unit, resulting in total ion beam intensities of $\sim 1.2\times 10^{-10}$ A. Instrumental mass bias was corrected by internal normalization to $^{98}Mo/^{96}Mo$ =

1.453173 using the exponential law. The Mo isotope data are reported in the ε-notation relative to the bracketing runs of the Alfa Aesar Mo solution standard. For all samples, the reported data represent the mean of repeat measurements. The stated uncertainties reflect 95% CI for samples with N>3 or the external reproducibility (2 s.d.) obtained from repeated analysis of BHVO-2 for samples with N≤3 [i.e. ±0.22 for $\varepsilon^{94}$Mo and ±0.15 for $\varepsilon^{95}$Mo; data reported in Budde et al. (2019)]. The accuracy and precision of the Mo isotope measurements were assessed by repeated analyses of BHVO-2, for which we obtained $\varepsilon^{i}$Mo values of ~0 throughout.

# 3 Results

## *3.1 Bulk meteorites*

### *3.1.1 Ti isotopes*

The Ti isotopic data obtained in this study for bulk meteorites and terrestrial rocks are provided in Tables 1–2. Consistent with results of prior studies, the bulk NC meteorites are characterized by negative and correlated $\varepsilon^{46}$Ti and $\varepsilon^{50}$Ti values (Fig. S1). With the exception of brachinites NWA 3151 and NWA 10637 and the brachinite-like achondrite GRA 06128, the samples show no resolved $^{48}$Ti isotope anomalies. Samples from a given group of NC meteorites usually have indistinguishable Ti isotope compositions, where ureilites have the lowest $\varepsilon^{46}$Ti and $\varepsilon^{50}$Ti values. Other achondrites analyzed in this study (i.e. acapulcoite-lodranites, brachinites, angrites, HEDs, main group pallasites, and mesosiderites) and the K chondrite have smaller Ti isotope anomalies ($\varepsilon^{50}$Ti ≈ –1.3), which are then followed by R chondrites and winonaites ($\varepsilon^{50}$Ti ≈ –0.4), and enstatite chondrites (EC) and aubrites ($\varepsilon^{50}$Ti ≈ 0 to –0.2). The ungrouped NC chondrites and achondrites of this study fall within this range of Ti isotope compositions and, importantly, fill some of the apparent gaps in the $\varepsilon^{50}$Ti range defined by the major meteorite groups (Fig. 1).

The CC meteorites of this study are characterized by positive and correlated $\varepsilon^{46}$Ti and $\varepsilon^{50}$Ti values, which is again consistent with results of prior studies (e.g., Trinquier et al., 2009). As for the NC meteorites, there are no clearly resolved $^{48}$Ti isotope anomalies. The CV, CM, and CL chondrites investigated here fall well within the ranges of previously reported Ti isotopic compositions for these groups (Table 4). The ungrouped chondrites are characterized by variable $\varepsilon^{46}$Ti and $\varepsilon^{50}$Ti values that nevertheless fall within the range of known Ti isotopic compositions defined by the major carbonaceous chondrite groups. The different Allende subsamples display variable Ti isotope compositions with a range of ~2 $\varepsilon^{50}$Ti, as expected from the different amounts of Ca-Al-rich inclusions (CAIs) contained in them. Finally, NWA 6926 displays only a moderate $^{50}$Ti excess, consistent with Ti isotopic data for other ungrouped carbonaceous achondrites (Sanborn et al., 2019; Williams et al., 2020).

As for the NC meteorites alone, the $\varepsilon^{50}$Ti and $\varepsilon^{46}$Ti values of all bulk meteorites of this study are linearly correlated with a slope of 5.7±0.1 (95% CI, $n$ = 66; Fig. 1), in agreement with the slope of 5.6±0.1 defined by a previous combined regression of NC meteorites, CC meteorites, and CAIs (Torrano et al., 2019).

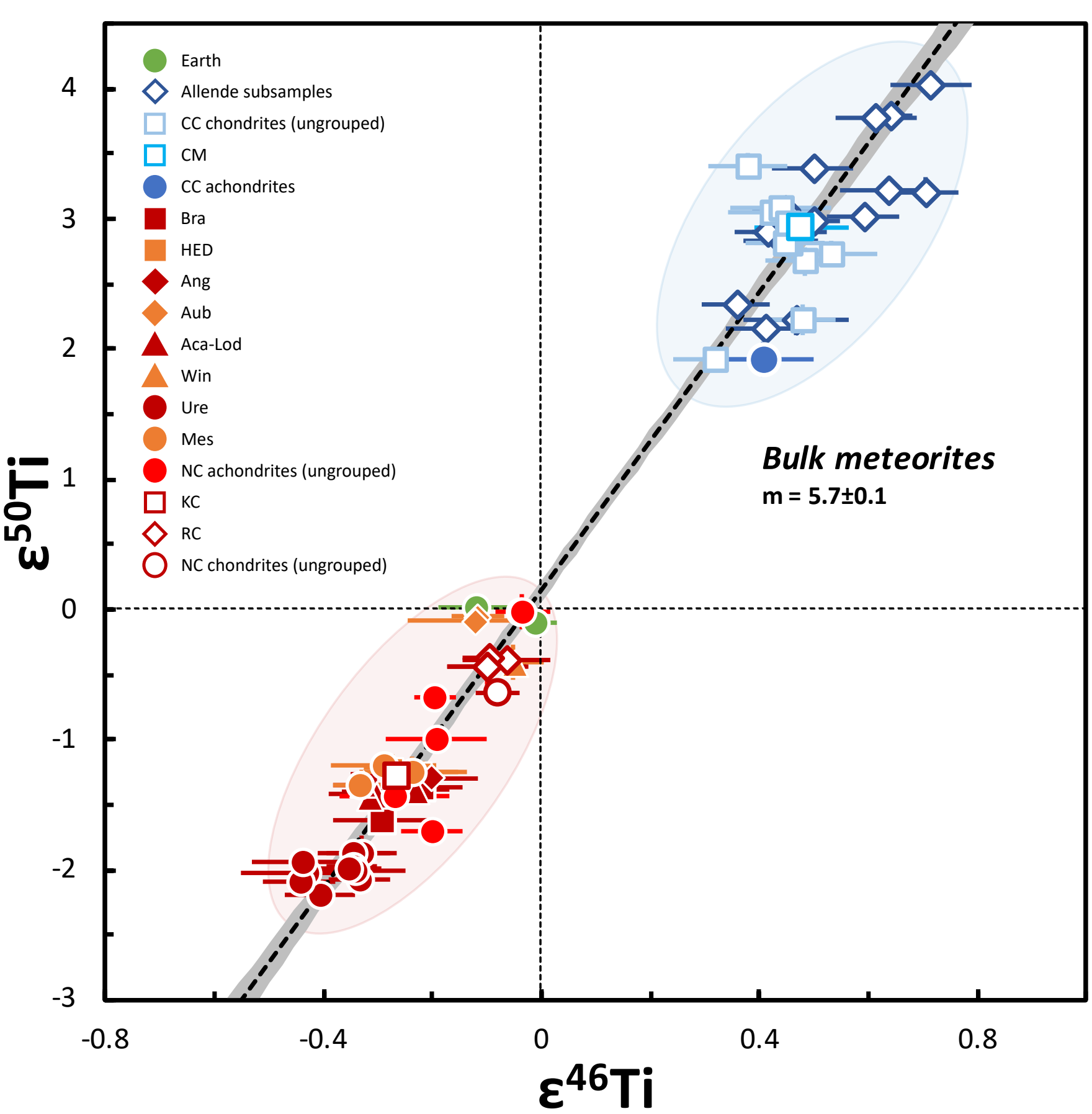

**Fig. 1:** Titanium isotope compositions of non-carbonaceous (NC; red) and carbonaceous (CC; blue) meteorites obtained in this study. Chondrites are shown with open, achondrites with filled symbols. The dashed line represents a regression of the combined NC and CC data and was calculated using IsoplotR (95% CI error envelope). Literature data is included in the shaded ellipses (see Table 4 for details and references). Abbreviations: Bra – Brachinites, HED – Howardites-Eucrites-Diogenites, Ang – Angrites, Aub – Aubrites, Aca-Lod – Acapulcoite-Lodranites, Win – Winonaites, Ure – Ureilites, Mes – Mesosiderites, KC – Kakangari chondrites, RC – Rumuruti chondrites.

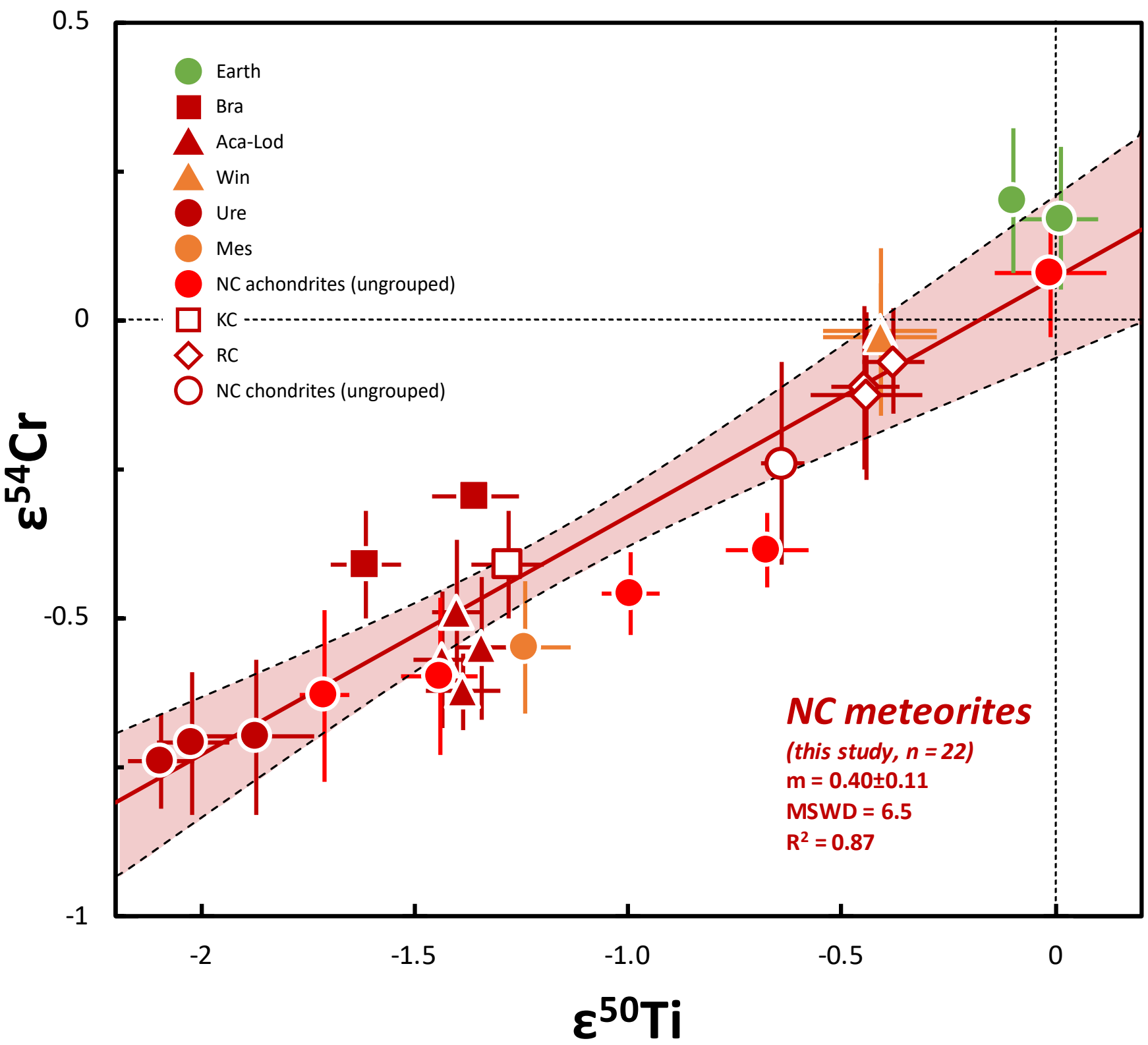

**Fig. 2:** $\varepsilon^{54}$Cr vs. $\varepsilon^{50}$Ti for the NC meteorites investigated in this study. Chondrites are shown with open, achondrites with filled symbols. The red line represents a regression of the data (excluding terrestrial samples) calculated using IsoplotR (95% CI error envelope). Note that our data are consistent with literature data (see Table 4 for details and references). Abbreviations: Bra – Brachinites, Aca-Lod – Acapulcoite-Lodranites, Win – Winonaites, Ure – Ureilites, Mes – Mesosiderites, KC – Kakangari chondrites, RC – Rumuruti chondrites.

### *3.1.2 Cr isotopes*

The Cr isotopic data for 22 of the bulk meteorite samples analyzed in this study are reported in Tables 1–2. The newly investigated samples fall within the range of Cr isotope anomalies previously reported for the major NC meteorite groups (e.g., Trinquier et al., 2007; Qin et al., 2010; Zhu et al., 2021) with $\varepsilon^{54}$Cr values between ca. –0.74 for ureilite MS-MU-20 and ~0.08 for enstatite achondrite NWA 2526 (Fig. 2 and Fig. S2). Two samples of this study (winonaite sample HaH 193 and lodranite NWA 7474) show evidence for more elevated $\varepsilon^{54}$Cr values induced by spallation reactions due to the interaction with galactic cosmic rays (GCR). After correction for this effect (see Supplementary Material), the Cr isotopic compositions of HaH 193 ($\varepsilon^{54}$Cr = –0.03±0.08) and NWA 7474 ($\varepsilon^{54}$Cr = –0.55±0.12) are consistent with $\varepsilon^{54}$Cr values previously reported for these groups (Goodrich et al., 2017; Li et al., 2018; Rüfenacht et al., 2023). Consistent with results of prior studies, $\varepsilon^{54}$Cr and $\varepsilon^{50}$Ti values of the NC meteorites of this study are linearly correlated (Fig. 2).

The NC meteorites of this study also exhibit variable $\varepsilon^{53}$Cr values, including very radiogenic compositions of 1.24±0.05 for the ungrouped achondrite GRA 06128. The radiogenic $^{53}$Cr variations in some of the investigated meteorites (e.g., winonaites) can be used to infer the timing of Mn-Cr fractionation in their parent bodies.

Finally, the only CC sample for which we present new Cr isotope data (NWA 6926) displays an elevated $\varepsilon^{54}$Cr value of 1.44±0.11, consistent with Cr isotopic data for other ungrouped carbonaceous achondrites (e.g., Sanborn et al., 2019; Williams et al., 2020).

#### *3.1.3 Mo isotopes*

The new Mo isotope data for four bulk NC meteorites and one bulk CC meteorite are reported in Tables 1–2 (and Table S2). Also given in these tables are the Mo isotope data for the samples for which Cr and Ti isotopes were analyzed in this study on digestion aliquots from earlier Mo studies (Worsham et al., 2017; Budde et al., 2019; Hopp et al., 2020). The new Mo isotope data for NC samples cover the entire range of Mo isotope anomalies reported for bulk NC meteorites (Fig. 3; Table 4). The CL chondrite NWA 13400 displays elevated $\varepsilon^{94}$Mo–$\varepsilon^{95}$Mo values, similar to CV and CO chondrites and consistent with its affinity to the CC reservoir.

Overall, the NC meteorites exhibit correlated $\varepsilon^{94}$Mo-$\varepsilon^{46}$Ti-$\varepsilon^{50}$Ti-$\varepsilon^{54}$Cr variations (Fig. 4), where in particular the chondrites (but also some iron meteorites; Fig. 4D) plot on well-defined correlation lines. Most of the achondrites also plot on or close to these correlation lines, but some samples plot off these lines.

### *3.2 Ordinary chondrite leachates*

The leachate fractions of the ordinary chondrites NWA 2458 and WSG 95300 display Ti isotopic variability ranging from negative $\varepsilon^{50}$Ti in L1 to more positive $\varepsilon^{50}$Ti values in later leaching steps (Table 3). For both meteorites the majority of Ti is released during leaching step L4 (e.g., ~80% for NWA 2458 and ~50% for WSG 95300; Fig. S3), which accordingly exhibits Ti isotopic compositions similar to bulk OCs ($\varepsilon^{50}$Ti = –0.62±0.04; Tables 3–4). The $\varepsilon^{46}$Ti and $\varepsilon^{50}$Ti values of the leachates are positively correlated, especially for WSG 95300 (Fig. 5A). Importantly, this correlation is steeper than that defined by bulk meteorites.

For comparison, the Mo isotopic compositions (from Budde et al., 2019) of the twelve OC leachate fractions investigated here for Ti are also provided in Table 3. Unlike for the bulk NC meteorites, there is no obvious correlation between $\varepsilon^{94}$Mo (or $\varepsilon^{95}$Mo) and $\varepsilon^{50}$Ti (Fig. 5B).

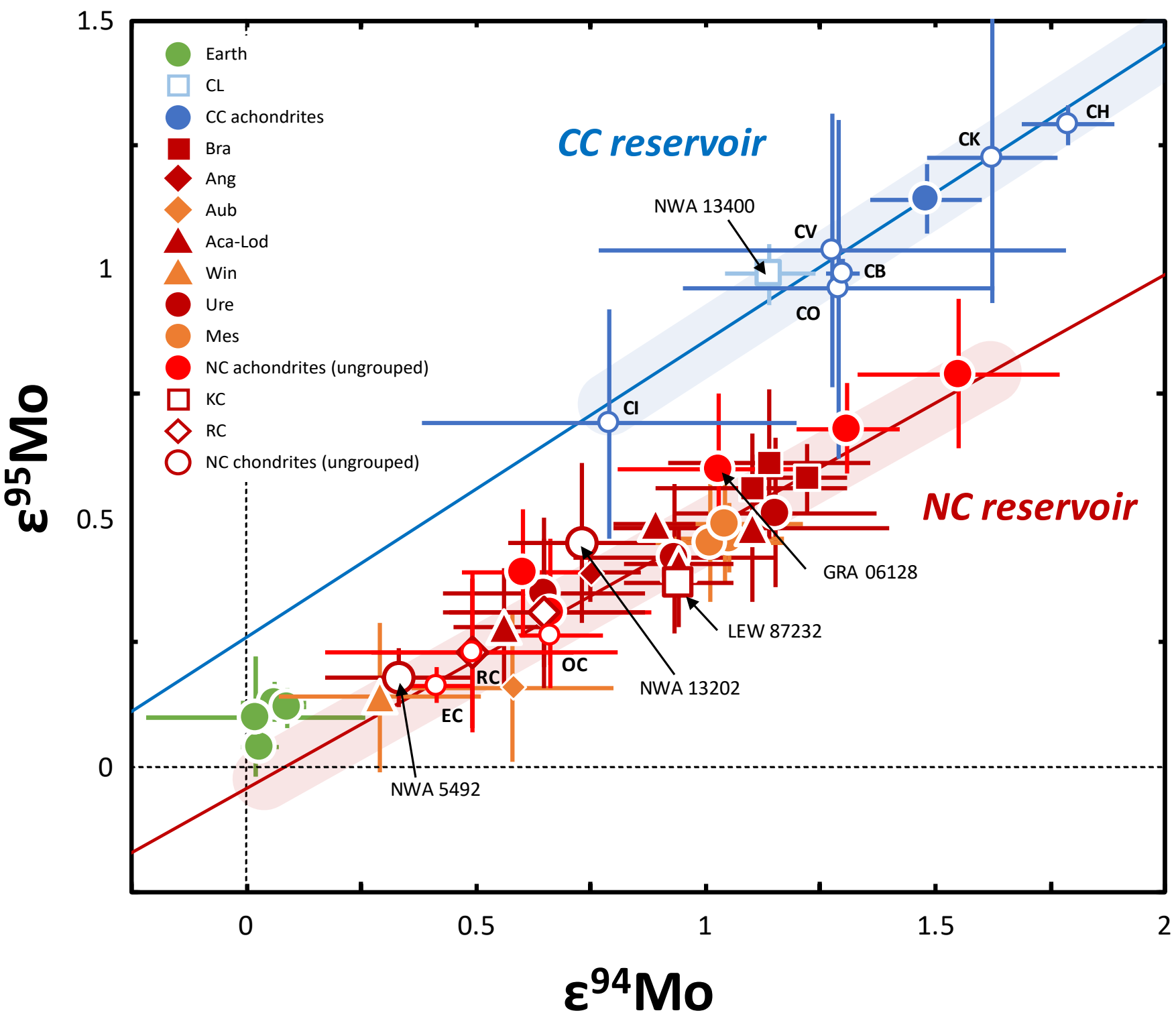

**Fig. 3:** Molybdenum isotope systematics of NC and CC meteorites. The five meteorites measured in this study (NWA 13400, GRA 06128, LEW 87232, NWA 13202, NWA 5492) are labelled. Mo isotope data from previous studies conducted on the same solution/powder aliquots as the Ti and Cr isotope measurements of this study are shown by the larger symbols, while the major NC and CC chondrite groups are shown for comparison with smaller symbols. Literature data is included in the shaded ellipses (see Table 4 for details and references). The slopes of the NC and CC lines are from Spitzer et al. (2025a) and Budde et al. (2019), respectively. Abbreviations: Bra – Brachinites, Ang – Angrites, Aub – Aubrites, Aca-Lod – Acapulcoite-Lodranites, Win – Winonaites, Ure – Ureilites, Mes – Mesosiderites, KC – Kakangari chondrites, RC – Rumuruti chondrites.

# 4 Discussion

## *4.1 Planetary isotope genetics*

### *4.1.1 NC chondrites*

*Kakangari chondrites:* Isotopic data for members of the K chondrite grouplet are sparse. Their petrologic, bulk chemical, and O isotope characteristics show similarities to several of the major chondrite groups (Weisberg et al., 1996). For example, while chondrules in K chondrites have O isotopic compositions similar to ECs, the bulk chondrites and matrix are more $^{16}$O-rich. This is due to the ubiquitous presence of AOA-like $^{16}$O-rich grains, a feature also seen in the matrix of carbonaceous chondrites (Weisberg et al., 1996; Nagashima et al., 2015). Despite these CC-like features, the Ti, Cr, and Mo isotope anomalies reported here for LEW 87232 demonstrate that the K chondrites belong to the NC meteorites. Importantly, their coupled Ti-Cr-Mo isotope signatures are more anomalous than those of other NC chondrite groups (Fig. 4, Table 4), and are more similar to those observed for brachinites, acapulcoites, and other NC achondrites. The H4 chondrite GRV 020043 has so far been the only chondritic sample showing large $^{50}$Ti–$^{54}$Cr deficits ($\varepsilon^{50}$Ti = –1.59±0.24 and $\varepsilon^{54}$Cr = –0.48±0.10; Li et al., 2018; Williams et al., 2020), but it is unclear as to whether this sample is a chondrite or rather a primitive achondrite related to the acapulcoites-lodranites (Li et al., 2018). Thus, at present the K chondrites represent the NC chondrites with the largest isotope anomalies and which substantially extend the range of isotope anomalies observed among NC chondrites (Fig. 4).

*Rumuruti chondrites:* The four investigated RC samples display only moderate $^{50}$Ti–$^{54}$Cr deficits and $^{94}$Mo excesses, in-between those of the ECs and OCs, and similar to the isotopic composition of Mars (Zhu et al., 2021a; Zhu et al., 2022; Burkhardt et al., 2021; Rüfenacht et al., 2023) (Fig. 4). This contrasts with the RC's endmember isotope signature for O (Bischoff et al., 2011) and implies that there is no simple relationship between nucleosynthetic isotope anomalies and mass-independent O isotopic variations among NC meteorites. This decoupling may reflect modifications of the O isotope composition by the interaction of the chondrite's components with nebular gas (Clayton, 2008) or the addition of $^{17,18}$O-enriched water to the precursors of the RCs, in line with their oxidized mineralogy (Bischoff et al., 2011).

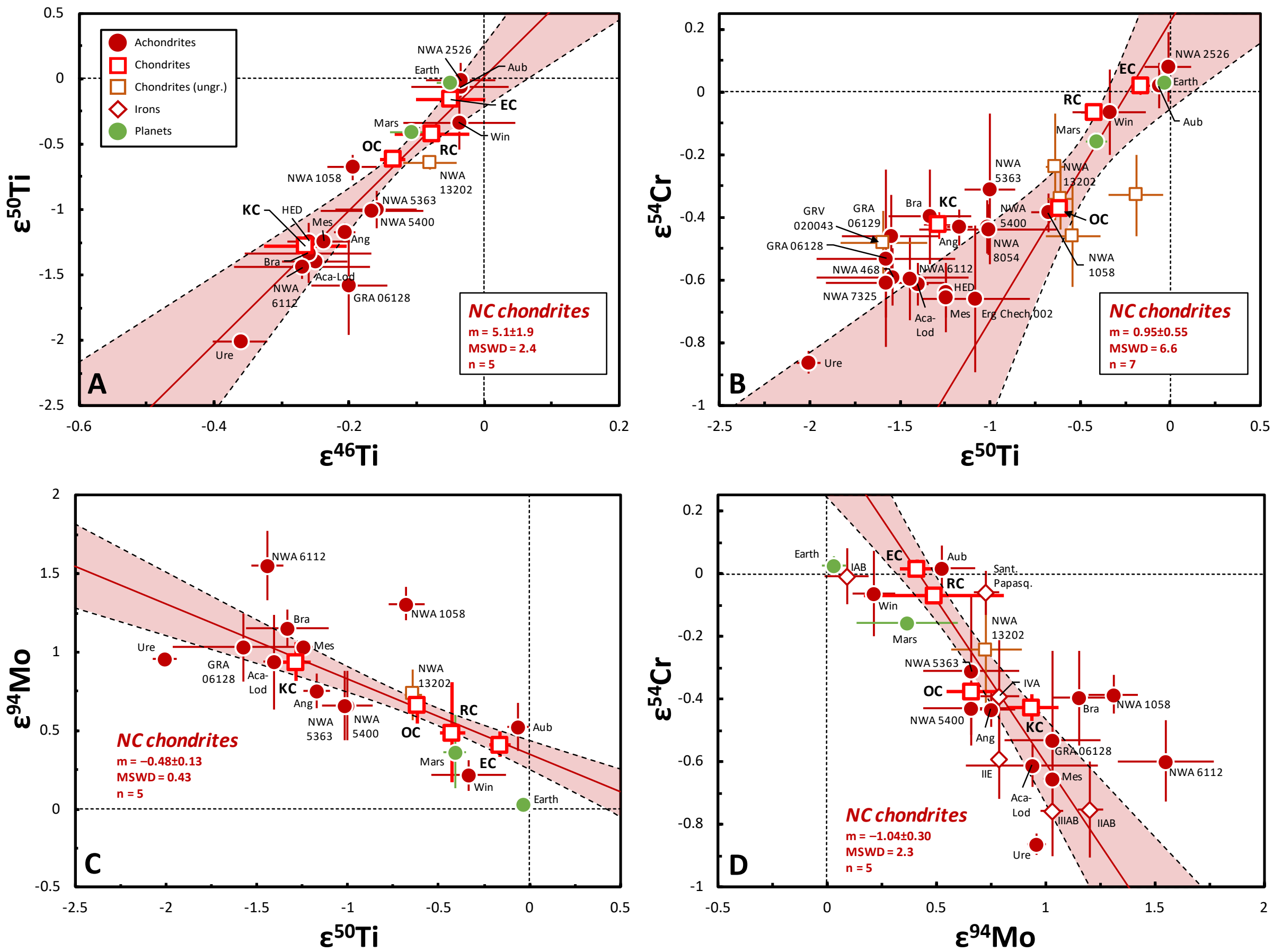

**Fig. 4:** Multi-element (Ti, Cr, Mo) isotope plots illustrating the correlated isotope variations among non-carbonaceous (NC) meteorites. In each panel, meteorite groups are shown as composite points averaging multiple samples within each group [combination of literature data and data from this study see Table 4; $\varepsilon^{54}$Cr data of NC irons as compiled in Tissot et al. (2025) and updated with the data from Yang et al. (2025a); $\varepsilon^{94}$Mo data of NC irons as compiled by Spitzer et al. (2025a)]. As such, each data point represents an individual meteorite parent body. Chondrites are shown by open, achondrites by filled symbols. Note that there is scatter beyond analytical uncertainty along these trends, regardless of which of the three elements are plotted against each other. Most of the scatter comes from the composition of (primitive) achondrites, while chondrites (and also most iron meteorites) plot along reasonably well-defined correlation lines. Red lines represent linear regressions of NC chondrites, were calculated using IsoplotR (95% CI error envelopes). Kakangari chondrites and GRV 020043 were not included in the regression in panel (B); see main text for details.

*Metal-rich chondrites:* NWA 5492, together with GRO 95551 and Sierra Gorda 009, has been assigned to the G chondrite (GC) grouplet (Weisberg et al., 2015; Ivanova et al., 2020). Unlike the metal-rich carbonaceous CB- and CH-type chondrites, G chondrites show mineralogical, petrological, as well as O, C, and N isotopic similarities to NC chondrites. The new Mo isotopic data of this study reveal an NC origin of metal in NWA 5492 most similar to ECs (Fig. 3), consistent with the $^{54}$Cr isotopic composition of the G chondrite GRO 95551 (Zhu et al., 2023). Due to their metal-rich nature and lack of matrix, it has been suggested that the GCs formed by a large-scale impact event analogous to the metal-rich CB/CH chondrites (Krot et al., 2005; Weisberg et al., 2015; Ivanova et al., 2020). Within this framework, the combined Mo and Cr isotopic data would indicate that the collision occurred between two NC bodies. Similarly, the metal-rich, ungrouped chondrite NWA 13202 displays petrologic properties similar to both CBs and OCs, and an impact origin has also been suggested for this chondrite (Jansen et al., 2019). The Ti, Cr, and Mo isotopic compositions measured for NWA 13202 in this study are indistinguishable from OCs, indicating an NC origin of NWA 13202 and—in case of an impact origin—of the colliding bodies that produced the NWA 13202 parent body.

### *4.1.2 CC chondrites*

The CL chondrite group has only recently been defined (Metzler et al., 2021) and nucleosynthetic isotope anomaly data for this group has so far been limited to Ti and Cr. The Mo isotopic composition of the CL chondrite NWA 13400 determined in this study confirms a CC origin of this group and is most similar to CV and CO chondrites (Fig. 3). This is consistent with the similar $^{54}$Cr compositions of CL and CV/CO chondrites. By contrast, the $^{50}$Ti composition of CL chondrites is different from CV chondrites, consistent with the $\varepsilon^{50}$Ti value measured here for NWA 13400 (and in Metzler et al., 2021). The two ungrouped carbonaceous chondrites DaG 055 (C3-ung) and DaG 430 (C3-ung) are probably paired (together with DaG 056) and show mineralogical and petrological similarities to both CL and CV chondrites (Scherer and Schultz, 2000). However, their Ti isotopic compositions as determined in this study suggest they are genetically related to CL rather than CV chondrites. As such, these samples may be the first identified type 3 samples of the CL chondrite group.

One problem with using Ti isotope variations in carbonaceous chondrites to assess genetic relationships is that $\varepsilon^{50}$Ti values may vary at the sampling scale owing to the heterogeneous distribution of CAIs. These have relatively large $^{50}$Ti excesses and owing to their elevated Ti abundance, exert a strong control on the Ti isotopic composition of a sample (e.g., Trinquier et al., 2009; Davis et al., 2018; Burkhardt et al., 2019; Torrano et al., 2019, 2023; Zhu et al., 2025). This is evident from the $\varepsilon^{50}$Ti variations observed among the subsamples of the Allende CV3 chondrite investigated in this study. As can be seen in Fig. 1 of Stracke et al. (2012), the subsamples C5, G2, and A2 include visible CAIs. Consistent with this, these subsamples are also characterized by the largest $^{50}$Ti excesses of up to $\varepsilon^{50}$Ti ≈ 4 together with the highest $TiO_2$ contents, which is a chemical proxy for the abundance of Ti-rich, refractory inclusions (especially CAIs; Fig. S4). By contrast, the other subsamples are mostly dominated by matrix and chondrules, and in particular subsamples A4, A6, and D4 show only moderate $^{50}$Ti excesses of $\varepsilon^{50}$Ti ≈ 2.2 and low $TiO_2$ contents. Together, these results imply that the Ti isotope compositions of individual CC chondrites obtained on small samples may not reflect the composition of the true bulk rock. As such it is important to recognize that both DaG 055 and DaG 430 exhibit similar $\varepsilon^{50}$Ti versus $TiO_2$ systematics as the Allende subsamples of this study (Fig. S4), introducing some uncertainty in assigning these samples to either the CL or CV chondrites.

#### *4.1.3 Achondrites*

*'Brachinite-like' achondrites:* GRA 06128, NWA 5400, and NWA 6112 display distinct $\varepsilon^{50}$Ti values, and only NWA 6112 is characterized by a Ti isotopic composition ($\varepsilon^{50}$Ti = –1.44 ± 0.09) typically observed for brachinites ($\varepsilon^{50}$Ti ≈ –1.3, Table 4). By contrast, GRA 06128 shows a lower and NWA 5400 a higher $\varepsilon^{50}$Ti than observed for brachinites. Likewise, the $\varepsilon^{54}$Cr of all three samples, especially of GRA 06128 and NWA 6112 ($\varepsilon^{54}$Cr = –0.63±0.14 and –0.60±0.13, respectively), are different from those observed for the brachinites investigated here (average $\varepsilon^{54}$Cr = –0.35±0.16). The $\varepsilon^{94}$Mo values of all three samples are also distinct from each other and cover almost the entire range of Mo isotope variations observed among NC meteorites (Fig. 3); only the $\varepsilon^{94}$Mo of GRA 06128 is similar to that of brachinites. Although some of this variability might be due to parent-body processes and terrestrial alteration (Yokoyama et al., 2019), the combined Ti-Cr-Mo anomalies suggest that the three 'brachinite-like' ungrouped achondrites sample distinct parent bodies.

*Northwest Africa 1058:* Based on petrologic and/or O isotopic similarities, this ungrouped achondrite has been genetically linked to acapulcoite-lodranites, diogenites, or winonaites (Eugster and Lorenzetti, 2005). However, while the $^{50}$Ti and $^{54}$Cr isotopic compositions of NWA 1058 determined in this study are distinct from these NC achondrites and more similar to those of ordinary chondrites, the Mo isotopic composition of NWA 1058 is distinct from the OCs but more similar to some NC achondrites (Fig. 3). Thus, NWA 1058 likely originates from a distinct parent body than the other achondrites. Of note, winonaites themselves display distinct Ti, Cr, and Mo isotope anomalies compared to acapulcoite-lodranites, and so although these primitive achondrites share some textural and petrological characteristics (Weisberg et al., 2006; Dhaliwal et al., 2017; Keil and McCoy, 2018), they likely originate from genetically distinct parent bodies.

*Northwest Africa 2526:* As an enstatite-rich achondrite, NWA 2526 has been interpreted as a partial melt residue from an EC-like parent body (Keil and Bischoff, 2008). Consistent with this, the Ti and Cr isotopic compositions of this sample are indistinguishable from ECs and aubrites (Zhang et al., 2012; Zhu et al., 2021c). Likewise, NWA 2526 exhibits an EC-like Ru isotope composition, although its Mo isotope anomalies are slightly distinct, possibly due to heterogeneous sampling or selective processing of *s*-process Mo carriers on the parent body (Hopp et al., 2020).

*Northwest Africa 6926:* The $^{50}$Ti and $^{54}$Cr excesses measured for this ungrouped achondrite are consistent with its CC heritage inferred from Mo isotopes (Hopp et al., 2020), and the Ti and Cr isotopic composition of its proposed paired meteorites NWA 6693 and NWA 6704 (Sanborn et al., 2019; Williams et al., 2020). Based on its O isotopic and bulk chemical composition NWA 6926 has been interpreted to represent an igneous cumulate from an unusually oxidized body distinct from other CC-type achondrites (Warren et al., 2013). In $^{54}$Cr–$^{50}$Ti isotope space, however, NWA 6926 does not seem to be different from other CC-type achondrites, suggesting that the more oxidized nature of this sample reflects processes on the parent body rather than genetic differences.

### *4.2 Isotope anomalies in ordinary chondrite leachates versus bulk NC meteorites*

The sequential digestion (leaching) of primitive chondrites is a powerful way of investigating the intrinsic, component-based isotopic variability of the disk at the time and place of the accretion of the chondrite parent bodies (Rotaru et al., 1992). Owing to their more primitive nature and thus higher abundance of presolar grains, this procedure has until now mostly been applied to carbonaceous chondrites, although some isotope data for leachates of NC chondrites also exist (Wang et al., 2011; Qin et al., 2011; Budde et al., 2019; Kadlag et al., 2021; Frossard et al., 2022, 2024). However, none of these prior studies on NC chondrites have investigated Ti isotope variations, and so the Ti isotopic data of this study are the first to allow assessing the origin and nature of Ti isotope variations among the presolar components of the NC reservoir.

The total ~5 $\varepsilon^{50}$Ti range among the OC leachates is smaller than the ~11 and ~15 $\varepsilon^{50}$Ti ranges observed among leachates of the CI chondrite Orgueil and the CM chondrite Murchison (Trinquier et al., 2009; Burkhardt et al., 2019), most likely reflecting a higher degree of isotopic homogenization of the original presolar grain population during thermal metamorphism on the OC parent bodies, as well as an overall lower matrix content of OCs (Alexander et al., 1990; Barosch et al., 2022). Nevertheless, a common feature of the NC and CC leachates is that their Ti isotope anomalies do not follow the slope defined by bulk meteorites (Fig. 5A). Instead, the NC leachates define a steeper slope with some excess scatter, suggesting the presence of multiple isotopically distinct Ti carriers (Trinquier et al., 2009; Steele and Boehnke, 2015; Burkhardt et al., 2019). Similarly, the combined $\varepsilon^{50}$Ti–$\varepsilon^{94}$Mo data for the OC leachates define no obvious correlation (Fig. 5B), unlike the bulk NC meteorites. While the lack of such a correlation for Ti and Mo may reflect the differential dissolution and redistribution of distinct isotopically anomalous carriers during parent body processes (e.g., Yokoyama et al., 2023b), these processes cannot account for the disparate slopes of isotope correlations for a single element, as observed in $\varepsilon^{46}$Ti–$\varepsilon^{50}$Ti space. Thus, consistent with inferences from CC leachates (Trinquier et al., 2009; Burkhardt et al., 2019), it appears that the presolar carriers still present in ordinary chondrites do not seem to be directly responsible for the correlated Ti isotope variations among the bulk NC meteorites.

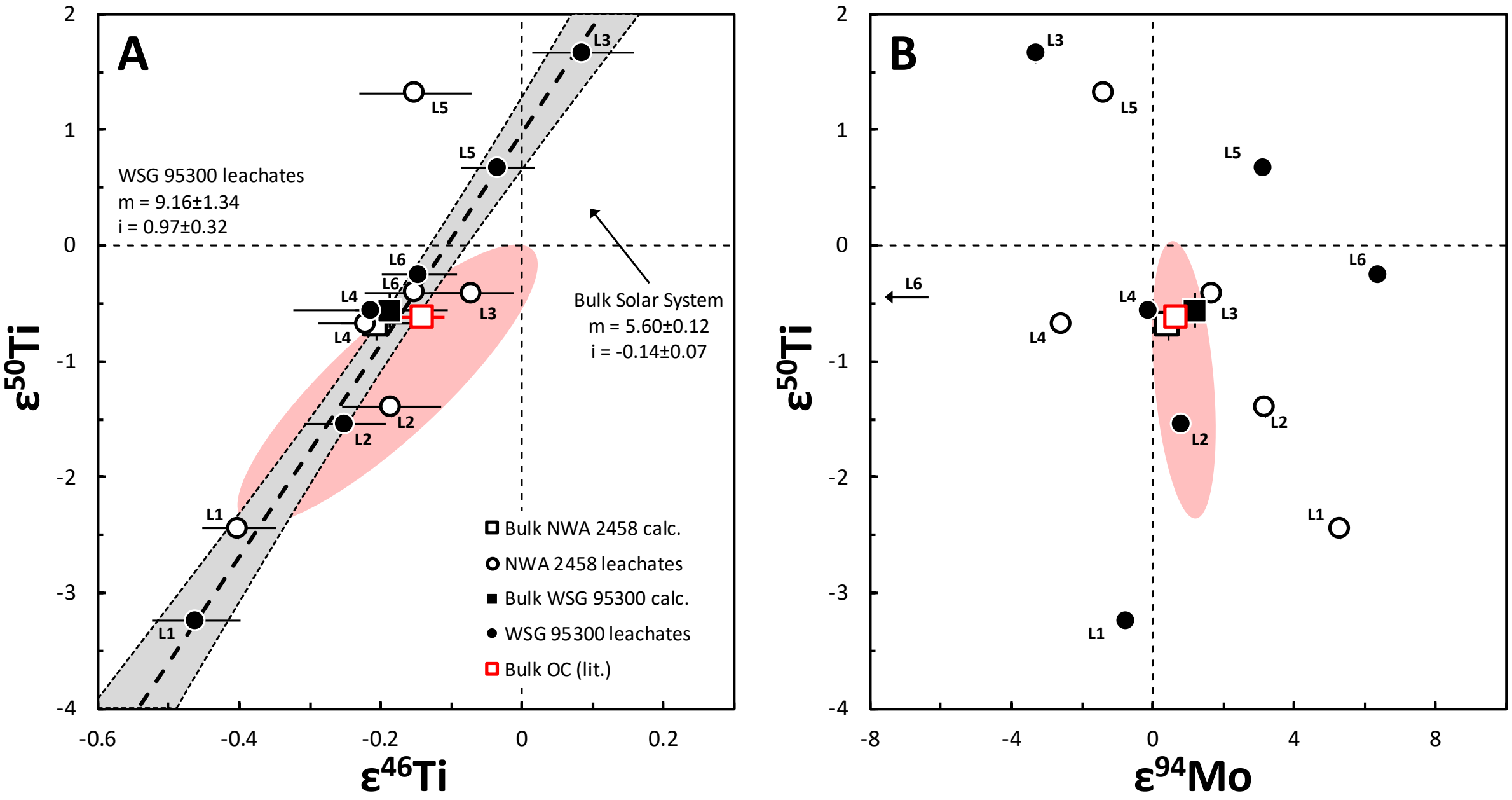

**Fig. 5:** Diagrams of (A) $\varepsilon^{50}$Ti vs. $\varepsilon^{46}$Ti and (B) $\varepsilon^{50}$Ti vs. $\varepsilon^{94}$Mo for acid leachates of the two unequilibrated ordinary chondrites NWA 2458 (L3.2; white circles) and WSG 95300 (H3.3; black circles). Uncertainties are 95% CI and are smaller than the symbol sizes in (B). The dashed black line and associated error envelope (95% CI) in (A) is a linear regression through the WSG-leachates and was calculated using IsoplotR. The solid black line represents the correlation defined by bulk meteorites, which is shallower than the slope defined by the leachates. Red ellipses show the overall range of isotopic variations among bulk NC meteorites.

### 4.3 *Origin of the NC isotope trend*

The NC meteorites display broadly correlated isotope anomalies for $^{46}$Ti, $^{50}$Ti, $^{54}$Cr, and $^{94}$Mo (Fig. 4), where some differentiated meteorites (e.g., ureilites, some iron meteorites) plot on one end of these trends, and enstatite chondrites, aubrites, and winonaites on the other end, close to the composition of the BSE. There is scatter beyond analytical uncertainty along these trends, regardless of which of the three elements are plotted against each other. Importantly though, most of the scatter comes from the composition of achondrites, while chondrites (and also most iron meteorites) plot along reasonably well-defined correlation lines. This is particularly evident in a plot of $\varepsilon^{50}$Ti vs. $\varepsilon^{94}$Mo (Fig. 4C), where the ECs, OCs, RCs, and KCs together with the ungrouped chondrite NWA 13202 plot along a single well-defined correlation line with no excess scatter (MSWD = 0.43; IsoplotR). While most achondrites also plot within uncertainty of this correlation line, several of them plot off the line, such as for example the winonaites, ureilites, and some ungrouped achondrites. A similar observation can be made from the $\varepsilon^{54}$Cr–$\varepsilon^{94}$Mo plot (Fig. 4D), where the NC chondrites also define a linear trend, albeit with more scatter (MSWD = 2.3; IsoplotR). Moreover, magmatic iron meteorites, which sample the metallic cores of differentiated planetesimals, also plot on this trend, while again some achondrites plot off this trend.

There are at least two explanations for why achondrites appear to preferentially plot off of the linear NC isotope trends defined by chondrites (and most iron meteorites). First, this may reflect selective processing of presolar carriers during partial melting on the parent bodies (Goderis et al., 2015). Although melting would be expected to homogenize any pre-existing isotopic heterogeneity in an object, it is conceivable that partial melting associated with rapid melt migration led to the selective removal of thermally labile and isotopically anomalous presolar host phases, thereby creating complementary isotope anomalies in the melting residues. This process was suggested to account for the strong enrichment of *s*-process Os in ureilites (Goderis et al., 2015) and for the decoupling of *s*-process signatures for Mo and Ru in brachinites and some primitive achondrites (Hopp et al., 2020). Importantly, all the samples deviating from the $\varepsilon^{50}$Ti–$\varepsilon^{94}$Mo and $\varepsilon^{54}$Cr–$\varepsilon^{94}$Mo trends defined by NC chondrites are primitive achondrites and as such derive from parent bodies that underwent only partial melting. These samples deviate from the NC trend mostly towards more *s*-process-depleted compositions (i.e. more positive $\varepsilon^{94}$Mo), and so if selective removal of presolar host phases during partial melting is responsible for this deviation, then *s*-process-rich Mo host phases must have been preferentially removed.

A second possibility is the heterogeneous distribution of chemically and isotopically distinct precursor materials among the NC parent bodies. For instance, isotopic variations among the carbonaceous chondrites are thought to reflect variable abundances of CAIs, chondrules, and matrix, all of which have distinct nucleosynthetic signatures (Alexander, 2019; Hellmann et al., 2023). NC chondrites are strongly depleted in CAIs (Bischoff and Keil, 1983; Fagan et al., 2000; Dunham et al., 2023), so the CAIs still present in them had no measurable effect on the isotopic composition of the bulk sample. However, some or most of the initial CAI population may have been reprocessed into chondrules and so can only be identified using the chondrule's chemical and isotopic composition. The $^{50}$Ti and $^{54}$Cr isotopic compositions of NC chondrules provides no evidence for the substantial incorporation of CAIs with $^{50}$Ti and $^{54}$Cr excesses (the typical composition of CAIs from carbonaceous chondrites) (e.g., Gerber et al., 2017). However, Na-Al-rich chondrules in ordinary chondrites are enriched in refractory elements and are characterized by $^{50}$Ti deficits, at one end of the NC trend and in some cases distinct from the $^{50}$Ti composition of their chondrite hosts (Ebert et al., 2018). Although the initial abundance of these refractory materials having $^{50}$Ti deficits is not known, it is conceivable that the heterogeneous distribution of these materials among different NC bodies produced some scatter along the NC trend. For instance, in the $\varepsilon^{50}$Ti vs. $\varepsilon^{54}$Cr plot (Fig. 4B), the K chondrites and the 'acapulco-chondrite' GRV 020043 deviate from the trend defined by the ECs, RCs, OCs, and other ungrouped NC chondrites in a manner as expected for a higher abundance of a refractory component having a $^{50}$Ti deficit. As observed for CAIs, such a refractory component would be strongly enriched in Ti but not in Cr, and so its addition may change the $\varepsilon^{50}$Ti of a sample without affecting its $\varepsilon^{54}$Cr. There are several achondrites (e.g., angrites, brachinites, ungrouped achondrites) having a similar $^{50}$Ti–$^{54}$Cr composition as the K chondrites and GRV 020043, and so one implication of this model would be that the parent bodies of these achondrites were also enriched in this refractory component. Together these observations imply that the heterogeneous distribution of chemically and isotopically distinct precursor components may be responsible for some of the scatter observed along the NC trend, and for the isotopic variations along this trend.

### 4.4 Comparison of NC chondrites and differentiated meteorites

The samples of this study include previously not, or just poorly, investigated meteorite groups (e.g., RC, KC, acapulcoites-lodranites, brachinites, ureilites, winonaites, mesosiderites), and so the new Ti, Cr, and Mo isotopic data for these samples allow a much more comprehensive assessment of the overall range of isotope anomalies in the NC reservoir for a diverse set of parent bodies that formed over an extended period of time. A key observation from the new data is that the nucleosynthetic isotope signatures of NC chondrites span almost the same range of compositions as the differentiated NC meteorites (Fig. 1–4). Within this framework, the new data for K chondrites are particularly important, as they considerably extend the range of $^{50}$Ti anomalies among NC chondrites towards the more anomalous compositions observed for some differentiated NC meteorites. Although there is considerable uncertainty in some of the accretion ages, there is broad agreement that differentiated and partially differentiated parent bodies formed within the first ~1.5 Ma of the solar system, i.e. during a time when the short-lived radionuclide $^{26}$Al was sufficiently abundant to induce parent body melting. This is consistent with accretion ages for iron meteorite and achondrite parent bodies inferred from Hf-W chronology of core formation in these bodies (Kleine et al., 2012; Kruijer et al., 2014; Touboul et al., 2015; Budde et al., 2015). Chondrites, by contrast, derive from parent bodies that did not melt and, therefore, almost certainly formed more than ~2 Ma after solar system formation, when too little $^{26}$Al remained to induced melting and differentiation. This is consistent with chondrite parent body accretion ages inferred from the ages of the youngest chondrules from a given chondrite (e.g., Rudraswami and Goswami, 2007; Rudraswami et al., 2008; Villeneuve et al., 2009; Kita and Ushikubo, 2012; Pape et al., 2019; Siron et al., 2021) and thermochronology of NC parent bodies (e.g., Trieloff et al., 2003; Kleine et al., 2008; Sugiura and Fujiya, 2014; Blackburn et al., 2017). We note, however, that unlike for other NC chondrites, there are no chronological data available for the K chondrites which would allow assessing their accretion age directly. Since these chondrites have the largest $^{50}$Ti deficits among the NC chondrites, the inference of a similar range of $^{50}$Ti compositions between differentiated and undifferentiated meteorites depends on the assumption that the K chondrites formed as late as other NC chondrites. Nevertheless, this assumption is reasonable, as available chronological data for chondrites generally indicate late accretion times and because there is no obvious reason why the K chondrites specifically should have formed earlier. We therefore conclude that despite their different accretion ages, early-formed differentiated meteorites and later-formed chondrites show a large overlap in their isotopic compositions (Fig. 4). The lack of any obvious temporal evolution of the NC isotopic composition implies that there was a significant dust-drift barrier isolating the NC and CC reservoirs for essentially the entire life-time of the disk. This is consistent with prior observations based primarily on the distinct Mo isotope signatures of NC and CC meteorites (e.g., Kleine et al., 2020; Spitzer et al., 2020).

The efficient isolation of the NC reservoir from inward-drifting CC dust implies that there was no significant late-stage replenishment of dust in the inner disk. This is problematic because any dust present in the inner disk is expected to have been lost on short timescales, either by accretion into an early generation of planetesimals or by drift into the Sun (e.g., Drążkowska et al., 2014; Birnstiel et al., 2016; Morbidelli et al., 2024). Thus, since there was no replenishment of dust from the outer disk, this raises the question from which dust the NC chondrites ultimately formed. One possibility is that this dust was produced by collisions among an earlier generation of planetesimals. However, the occurrence of presolar grains and the presence of isotope heterogeneities among chondrules from primitive NC chondrites (e.g., Ebert et al., 2018; Budde et al., 2019; Frossard et al., 2024; Marrocchi et al., 2024) imply that the NC chondrites formed from primitive dust that has escaped prior homogenization. This makes it unlikely that NC chondrites formed from secondary dust produced by collisions among differentiated planetesimals. Instead, this dust would have to derive from an early generation of planetesimals that owing to their small size escaped melting and differentiation. While such planetesimals may have existed, it is unclear if they carried sufficient mass to be the precursors of the NC chondrites.

Another problem with producing NC chondrites from collisionally produced dust is that collisions involve mixing among bodies having distinct compositions, and so they would be expected to result in an overall smaller range of compositions as observed among the colliding bodies. However, as observed in this study and noted above, NC chondrites exhibit a similar range of $^{50}$Ti isotopic compositions as the earlier-formed differentiated NC meteorites. It thus appears that both early- and late-formed NC planetesimals formed from the same long-lived primordial population of dust. This in turn implies the existence of substructures in the inner disk that hampered complete loss of any NC dust that was not incorporated into the earlier generation of NC planetesimals. Similar to what has been proposed for the carbonaceous chondrites (Hellmann et al., 2023), such substructures may have also led to different efficiencies with which chemically and isotopically distinct dust components were accreted into planetesimals, thereby contributing to the observed range of isotopic compositions among NC meteorites and the scatter observed along the NC trend.

### *4.5 Are there isotopic gaps within the NC reservoir?*

Based on a smaller $^{50}$Ti isotope data set than used in the present study, Rüfenacht et al., (2023) argued that there are compositional gaps in the NC reservoir, producing distinct isotopic clusters. Specifically, these authors identified two clusters which they termed 'Earth-Mars-like' (with $\varepsilon^{50}$Ti between 0 and –0.6), and Vesta-like ($\varepsilon^{50}$Ti between –1 and –1.5), and noted that there may be a third, 'ureilite-like' cluster characterized by an $\varepsilon^{50}$Ti of –2. These clusters were inferred to represent distinct regions of the NC reservoir that were spatially isolated by substructures in the inner disk. However, as is evident in particular from the $\varepsilon^{50}$Ti–$\varepsilon^{46}$Ti (Fig. 1) and $\varepsilon^{50}$Ti–$\varepsilon^{54}$Cr plots (Fig. 2 and 4B), some of the samples of the present study appear to fill these gaps and exhibit isotopic compositions in between these apparent clusters. Thus, although compositional clusters exist, these more likely reflect non-representative sampling of an initially more continuous isotopic trend among NC planetesimals instead of planetesimal formation in spatially isolated regions of the inner disk. This does not mean that substructures did not exist in the inner disk and, as we argued above, they in fact appear to be needed for preserving late-stage dust in the inner disk and to facilitate the formation of NC chondrite parent bodies. However, our data suggest that these substructures did not produce spatially separated sub-reservoirs in the disk, but rather facilitated the preservation of a long-lived dust reservoir from which NC planetesimals accreted over an extended period of time. This scenario can account for the large isotopic variability observed among early-accreted, (fully) differentiated NC achondrites, and the similarly large variability among later-accreted, undifferentiated NC chondrites, as they solely sample distinct source regions with variable mixing ratios between isotopically distinct endmember compositions.

## 5 Conclusions

The new Ti, Cr, and Mo isotope data for previously not, or only rarely investigated bulk meteorites, combined with data from prior studies, provide genetic information for 17 distinct NC meteorite parent bodies. This allows a more comprehensive assessment of the origin of isotopic variations within the NC reservoir than previously possible. Consistent with prior results, our data reveal correlated Ti, Cr, and Mo isotope variations among NC meteorites, but also show that the Mo-Ti and $^{46}$Ti-$^{50}$Ti isotope variations among bulk NC meteorites differ from those seen in acid leachates from primitive ordinary chondrites. This indicates that the presolar components still present in ordinary chondrites are not directly responsible for the observed isotope variations, and implies that thermal processing of these presolar carriers in the disk was not the main cause for the NC isotope variability. Unlike in some prior proposals, this isotope variability also cannot reflect a continuous temporal isotopic evolution of the inner disk by the addition of CC-like dust from the outer disk, because early- and late-formed NC bodies (i.e., the parent bodies of differentiated and chondritic NC meteorites) display a similar range in isotopic compositions.

A key observation from the new data of this study is that the isotopic variations along the NC trend display more scatter than previously recognized. Some of this scatter may reflect selective processing and partial loss of presolar carriers during partial melting and rapid melt migration on their parent bodies. However, the heterogeneous distribution of chemically and isotopically distinct precursor materials among the NC parent bodies appears to provide the most plausible explanation for at least some of this scatter, and most notably for the $^{50}$Ti-$^{54}$Cr isotope variations. As shown in prior studies, the carbonaceous chondrites likely also have formed by heterogeneous mixing among genetically diverse materials (e.g., Alexander, 2019; Hellmann et al., 2023), albeit with different isotopic compositions than for the components of the NC meteorites. Such mixing processes, therefore, appear to have been important for generating isotope variations within both the NC and CC reservoirs. Producing the NC isotope variability in this manner requires the presence of substructures in the inner disk (i.e. the NC reservoir) to facilitate fractionation of distinct dust components according to their aerodynamic properties. Substructures in the inner disk have previously been inferred based on apparent gaps in the Ti isotope composition among NC meteorites (Rüfenacht et al., 2023), but the data of the present study show that these compositional gaps more likely reflect unrepresentative sampling of planetesimals from an initially continuous isotopic trend. Thus, rather than producing compositional gaps, substructures instead appear to have been important for the preservation of a long-lived dust reservoir in the inner disk from which NC planetesimals accreted over an extended period of time. This scenario can also account for the similar range of isotopic compositions of early- and late-formed planetesimals in the NC reservoir.

**Acknowledgments**

We gratefully acknowledge NASA, the Institute of Meteoritics (University of New Mexico), and Addi Bischoff (University of Münster) for providing meteorite samples for this study as well as U. Heitmann for sample preparation. We also thank Ke Zhu and three anonymous reviewers for their constructive comments and Liping Qin for efficient editorial handling. This work was funded by the Deutsche Forschungsgemeinschaft (DFG, German Research Foundation) – Project-ID 263649064 – TRR 170 and the European Research Council Advanced Grant HolyEarth (grant no. 101019380).

**Data Availability**

Data are available through Mendeley: 10.17632/pbdbzbsrvg.1.

**Appendix A. Supplementary Material**

Supplementary material associated with this article includes an extended discussion on the correction of CRE effects on the Cr isotopic composition as well as supplementary tables and figures.

## References

Alexander, C.M.O., 2019. Quantitative models for the elemental and isotopic fractionations in chondrites: The carbonaceous chondrites. Geochim. Cosmochim. Acta 254, 277–309.

Alexander, C.M.O., Arden, J.W., Ash, R.D., Pillinger, C.T., 1990. Presolar components in the ordinary chondrites. Earth Planet. Sci. Lett. 99, 220–229.

Anand, A., Kruttasch, P.M., Mezger, K., 2022. $^{53}$Mn-$^{53}$Cr chronology and ε$^{54}$Cr-Δ$^{17}$O genealogy of Erg Chech 002: The oldest andesite in the solar system. Meteorit. Planet. Sci. 57, 2003–2016.

Barosch, J., Nittler, L.R., Wang, J., Dobrică, E., Brearley, A.J., Hezel, D.C., Alexander, C.M.O., 2022. Presolar O- and C-anomalous grains in unequilibrated ordinary chondrite matrices. Geochim. Cosmochim. Acta 335, 169–182.

Bermingham, K.R., Füri, E., Lodders, K., Marty, B., 2020. The NC-CC Isotope Dichotomy: Implications for the Chemical and Isotopic Evolution of the Early Solar System. Space Sci. Rev. 216.

Bermingham, K.R., Worsham, E.A., Walker, R.J., 2018. New insights into Mo and Ru isotope variation in the nebula and terrestrial planet accretionary genetics. Earth Planet. Sci. Lett. 487, 221–229.

Birnstiel, T., Fang, M., Johansen, A., 2016. Dust Evolution and the Formation of Planetesimals. Space Sci. Rev. 205, 41–75.

Bischoff, A., Barrat, J.-A., Bauer, K., Burkhardt, C., Busemann, H., Ebert, S., Gonsior, M., Hakenmüller, J., Haloda, J., Harries, D., Heinlein, D., Hiesinger, H., Hochleitner, R., Hoffmann, V., Kaliwoda, M., Laubenstein, M., Maden, C., Meier, M.M.M., Morlok, A., Pack, A., Ruf, A., Schmitt-Kopplin, P., Schönbächler, M., Steele, R.C.J., Spurný, P., Wimmer, K., 2017. The Stubenberg meteorite-An LL6 chondrite fragmental breccia recovered soon after precise prediction of the strewn field. Meteorit. Planet. Sci. 52, 1683–1703.

Bischoff, A., Barrat, J.-A., Berndt, J., Borovicka, J., Burkhardt, C., Busemann, H., Hakenmüller, J., Heinlein, D., Hertzog, J., Kaiser, J., Maden, C., Meier, M.M.M., Morino, P., Pack, A., Patzek, M., Reitze, M.P., Rüfenacht, M., Schmitt-Kopplin, P., Schönbächler, M., Spurný, P., Weber, I., Wimmer, K., Zikmund, T., 2019. The Renchen L5-6 chondrite breccia – The first confirmed meteorite fall from Baden-Württemberg (Germany). Geochemistry 79, 125525.

Bischoff, A., Keil, K., 1983. Ca–Al-rich chondrules and inclusions in ordinary chondrites. Nature 303, 588–592.

Bischoff, A., Vogel, N., Roszjar, J., 2011. The Rumuruti chondrite group. Geochemistry 71, 101–133.

Blackburn, T., Alexander, C.M.O., Carlson, R., Elkins-Tanton, L.T., 2017. The accretion and impact history of the ordinary chondrite parent bodies. Geochim. Cosmochim. Acta 200, 201–217.

Budde, G., Burkhardt, C., Brennecka, G.A., Fischer-Gödde, M., Kruijer, T.S., Kleine, T., 2016. Molybdenum isotopic evidence for the origin of chondrules and a distinct genetic heritage of carbonaceous and non-carbonaceous meteorites. Earth Planet. Sci. Lett. 454, 293–303.

Budde, G., Burkhardt, C., Kleine, T., 2019. Molybdenum isotopic evidence for the late accretion of outer Solar System material to Earth. Nat. Astron. 3, 736–741.

Budde, G., Kruijer, T.S., Fischer-Gödde, M., Irving, A.J., Kleine, T., 2015. Planetesimal differentiation revealed by the Hf–W systematics of ureilites. Earth Planet. Sci. Lett. 430, 316–325.

Budde, G., Kruijer, T.S., Kleine, T., 2018. Hf-W chronology of CR chondrites: Implications for the timescales of chondrule formation and the distribution of $^{26}$Al in the solar nebula. Geochim. Cosmochim. Acta 222, 284–304.

Burkhardt, C., Dauphas, N., Hans, U., Bourdon, B., Kleine, T., 2019. Elemental and isotopic variability in solar system materials by mixing and processing of primordial disk reservoirs. Geochim. Cosmochim. Acta 261, 145–170.

Burkhardt, C., Dauphas, N., Tang, H., Fischer-Gödde, M., Qin, L., Chen, J.H., Rout, S.S., Pack, A., Heck, P.R., Papanastassiou, D.A., 2017. In search of the Earth-forming reservoir: Mineralogical, chemical, and isotopic characterizations of the ungrouped achondrite NWA 5363/NWA 5400 and selected chondrites. Meteorit. Planet. Sci. 52, 807–826.

Burkhardt, C., Kleine, T., Oberli, F., Pack, A., Bourdon, B., Wieler, R., 2011. Molybdenum isotope anomalies in meteorites: Constraints on solar nebula evolution and origin of the Earth. Earth Planet. Sci. Lett. 312, 390–400.

Burkhardt, C., Spitzer, F., Morbidelli, A., Budde, G., Render, J.H., Kruijer, T.S., Kleine, T., 2021. Terrestrial planet formation from lost inner solar system material. Sci. Adv. 7, eabj7601.

Clayton, R.N., 2008. Oxygen Isotopes in the Early Solar System -- A Historical Perspective. Rev. Mineral. Geochem. 68, 5–14.

Davis, A.M., Zhang, J., Greber, N.D., Hu, J., Tissot, F.L.H., Dauphas, N., 2018. Titanium isotopes and rare earth patterns in CAIs: Evidence for thermal processing and gas-dust decoupling in the protoplanetary disk. Geochim. Cosmochim. Acta 221, 275–295.

Day, J.M.D., Walker, R.J., Ash, R.D., Liu, Y., Rumble, D., Irving, A.J., Goodrich, C.A., Tait, K., McDonough, W.F., Taylor, L.A., 2012. Origin of felsic achondrites Graves Nunataks 06128 and 06129, and ultramafic brachinites and brachinite-like achondrites by partial melting of volatile-rich primitive parent bodies. Geochim. Cosmochim. Acta 81, 94–128.

Dhaliwal, J.K., Day, J.M.D., Corder, C.A., Tait, K.T., Marti, K., Assayag, N., Cartigny, P., Rumble, D., Taylor, L.A., 2017. Early metal-silicate differentiation during planetesimal formation revealed by acapulcoite and lodranite meteorites. Geochim. Cosmochim. Acta 216, 115–140.

Drążkowska, J., Windmark⋆, F., Dullemond, C.P., 2014. Modeling dust growth in protoplanetary disks: The breakthrough case. Astron. Astrophys. 567, A38.

Dunham, E.T., Sheikh, A., Opara, D., Matsuda, N., Liu, M. -C., McKeegan, K.D., 2023. Calcium–aluminum-rich inclusions in non-carbonaceous chondrites: Abundances, sizes, and mineralogy. Meteorit. Planet. Sci. 58, 643–671.

Ebert, S., Render, J., Brennecka, G.A., Burkhardt, C., Bischoff, A., Gerber, S., Kleine, T., 2018. Ti isotopic evidence for a non-CAI refractory component in the inner Solar System. Earth Planet. Sci. Lett. 498, 257–265.

Ek, M., Hunt, A.C., Lugaro, M., Schönbächler, M., 2020. The origin of *s*-process isotope heterogeneity in the solar protoplanetary disk. Nat. Astron. 4, 273–281.

Eugster, O., Lorenzetti, S., 2005. Cosmic-ray exposure ages of four acapulcoites and two differentiated achondrites and evidence for a two-layer structure of the acapulcoite/lodranite parent asteroid. Geochim. Cosmochim. Acta 69, 2675–2685.

Fagan, T.J., Krot, A.N., Keil, K., 2000. Calcium-aluminum-rich inclusions in enstatite chondrites (I): Mineralogy and textures. Meteorit. Planet. Sci. 35, 771–781.

Frossard, P., Bonnand, P., Boyet, M., Bouvier, A., 2024. Role of redox conditions and thermal metamorphism in the preservation of Cr isotopic anomalies in components of non-carbonaceous chondrites. Geochim. Cosmochim. Acta 367, 165–178.

Frossard, P., Israel, C., Bouvier, A., Boyet, M., 2022. Earth's composition was modified by collisional erosion. Science 377, 1529–1532.

Gerber, S., Burkhardt, C., Budde, G., Metzler, K., Kleine, T., 2017. Mixing and Transport of Dust in the Early Solar Nebula as Inferred from Titanium Isotope Variations among Chondrules. Astrophys. J. 841, L17.

Goderis, S., Brandon, A.D., Mayer, B., Humayun, M., 2015. *s*-Process Os isotope enrichment in ureilites by planetary processing. Earth Planet. Sci. Lett. 431, 110–118.

Goodrich, C.A., Kita, N.T., Yin, Q.-Z., Sanborn, M.E., Williams, C.D., Nakashima, D., Lane, M.D., Boyle, S., 2017. Petrogenesis and provenance of ungrouped achondrite Northwest Africa 7325 from petrology, trace elements, oxygen, chromium and titanium isotopes, and mid-IR spectroscopy. Geochim. Cosmochim. Acta 203, 381–403.

Göpel, C., Birck, J.-L., Galy, A., Barrat, J.-A., Zanda, B., 2015. Mn–Cr systematics in primitive meteorites: Insights from mineral separation and partial dissolution. Geochim. Cosmochim. Acta 156, 1–24.

Gurrutxaga, N., Drążkowska, J., Vaikundaraman, V., Kleine, T., 2026. Carbonaceous Chondrites Provide Evidence for Late-stage Planetesimal Formation in a Pressure Bump. ApJ 1003, 132.

Hasegawa, H., Mikouchi, T., Yamaguchi, A., Yasutake, M., Greenwood, R.C., Franchi, I.A., 2019. Petrological, petrofabric, and oxygen isotopic study of five ungrouped meteorites related to brachinites. Meteorit. Planet. Sci. 54, 752–767.

van Helden, G., Tielens, A.G.G.M., van Heijnsbergen, D., Duncan, M.A., Hony, S., Waters, L.B.F.M., Meijer, G., 2000. Titanium Carbide Nanocrystals in Circumstellar Environments. Science 288, 313–316.

Hellmann, J.L., Schneider, J.M., Wölfer, E., Drążkowska, J., Jansen, C.A., Hopp, T., Burkhardt, C., Kleine, T., 2023. Origin of Isotopic Diversity among Carbonaceous Chondrites. Astrophys. J. Lett. 946, L34.

Hopp, T., Budde, G., Kleine, T., 2020. Heterogeneous accretion of Earth inferred from Mo-Ru isotope systematics. Earth Planet. Sci. Lett. 534, 116065.

Ivanova, M.A., Lorenz, C.A., Humayun, M., Corrigan, C.M., Ludwig, T., Trieloff, M., Righter, K., Franchi, I.A., Verchovsky, A.B., Korochantseva, E.V., Kozlov, V.V., Teplyakova, S.N., Korochantsev, A.V., Grokhovsky, V.I., 2020. Sierra Gorda 009: A new member of the metal-rich G chondrites grouplet. Meteorit. Planet. Sci. 55.

Jansen, C.A., Brenker, F.E., Zipfel, J., Pack, A., Labenne, L., Nagashima, K., Krot, A.N., Bizzarro, M., Schiller, M., 2019. Mineralogy, petrology, and oxygen isotopic composition of Northwest Africa 12379, metal-rich chondrite with affinity to ordinary chondrites. Geochemistry 79, 125537.

Jansen, C.A., Burkhardt, C., Marrocchi, Y., Schneider, J.M., Wölfer, E., Kleine, T., 2024. Condensate evolution in the solar nebula inferred from combined Cr, Ti, and O isotope analyses of amoeboid olivine aggregates. Earth Planet. Sci. Lett. 627, 118567.

Jenniskens, P., Rubin, A.E., Yin, Q., Sears, D.W.G., Sandford, S.A., Zolensky, M.E., Krot, A.N., Blair, L., Kane, D., Utas, J., Verish, R., Friedrich, J.M., Wimpenny, J., Eppich, G.R., Ziegler, K., Verosub, K.L., Rowland, D.J., Albers, J., Gural, P.S., Grigsby, B., Fries, M.D., Matson, R., Johnston, M., Silber, E., Brown, P., Yamakawa, A., Sanborn, M.E., Laubenstein, M., Welten, K.C., Nishiizumi, K., Meier, M.M.M., Busemann, H., Clay, P., Caffee, M.W., Schmitt-Kopplin, P., Hertkorn, N., Glavin, D.P., Callahan, M.P., Dworkin, J.P., Wu, Q., Zare, R.N., Grady, M., Verchovsky, S., Emel'Yanenko, V., Naroenkov, S., Clark, D.L., Girten, B., Worden, P.S., (The Novato Meteorite Consortium), 2014. Fall, recovery, and characterization of the Novato L6 chondrite breccia. Meteorit. Planet. Sci. 49, 1388–1425.

Kadlag, Y., Becker, H., Harbott, A., 2019. Cr isotopes in physically separated components of the Allende CV3 and Murchison CM2 chondrites: Implications for isotopic heterogeneity in the solar nebula and parent body processes. Meteorit. Planet. Sci. 54, 2116–2131.

Kadlag, Y., Hirtz, J., Becker, H., Leya, I., Mezger, K., 2021. Early solar irradiation as a source of the inner solar system chromium isotopic heterogeneity. Meteorit. Planet. Sci. 56, 2083–2102.

Keil, K., Bischoff, A., 2008. Northwest Africa 2526: A partial melt residue of enstatite chondrite parentage. Meteorit. Planet. Sci. 43, 1233–1240.

Keil, K., McCoy, T.J., 2018. Acapulcoite-lodranite meteorites: Ultramafic asteroidal partial melt residues. Geochemistry 78, 153–203.

Kita, N.T., Ushikubo, T., 2012. Evolution of protoplanetary disk inferred from $^{26}$ Al chronology of individual chondrules. Meteorit. Planet. Sci. 47, 1108–1119.

Kleine, T., Budde, G., Burkhardt, C., Kruijer, T.S., Worsham, E.A., Morbidelli, A., Nimmo, F., 2020. The Non-carbonaceous–Carbonaceous Meteorite Dichotomy. Space Sci. Rev. 216, 55.

Kleine, T., Hans, U., Irving, A.J., Bourdon, B., 2012. Chronology of the angrite parent body and implications for core formation in protoplanets. Geochim. Cosmochim. Acta 84, 186–203.

Kleine, T., Nimmo, F., 2025. Origin of the Earth. Treatise on Geochemistry, Third edition. Elsevier, pp. 325–381.

Kleine, T., Touboul, M., Van Orman, J.A., Bourdon, B., Maden, C., Mezger, K., Halliday, A.N., 2008. Hf–W thermochronometry: Closure temperature and constraints on the accretion and cooling history of the H chondrite parent body. Earth Planet. Sci. Lett. 270, 106–118.

Krestianinov, E., Amelin, Y., Yin, Q.-Z., Cary, P., Huyskens, M.H., Miller, A., Dey, S., Hibiya, Y., Tang, H., Young, E.D., Pack, A., Di Rocco, T., 2023. Igneous meteorites suggest Aluminium-26 heterogeneity in the early Solar Nebula. Nat. Commun. 14.

Krot, A.N., Amelin, Y., Cassen, P., Meibom, A., 2005. Young chondrules in CB chondrites from a giant impact in the early Solar System. Nature 436, 989–992.

Kruijer, Touboul, M., Fischer-Godde, M., Bermingham, K.R., Walker, R.J., Kleine, T., 2014. Protracted core formation and rapid accretion of protoplanets. Science 344, 1150–1154.

Kruijer, T.S., Borg, L.E., Wimpenny, J., Sio, C.K., 2020. Onset of magma ocean solidification on Mars inferred from Mn-Cr chronometry. Earth Planet. Sci. Lett. 542, 116315.

Kruttasch, P.M., Anand, A., Warren, P.H., Ma, C., Mezger, K., 2024. 53Mn-53Cr chronometry of ureilites: Implications for the timing of parent body accretion, differentiation and secondary reduction. Geochim. Cosmochim. Acta 383, 108–119.

Larsen, K.K., Trinquier, A., Paton, C., Schiller, M., Wielandt, D., Ivanova, M.A., Connelly, J.N., Nordlund, Å., Krot, A.N., Bizzarro, M., 2011. Evidence for magnesium isotope heterogeneity in the solar protoplanetary disk. Astrophys. J. 735, L37.

Li, S., Yin, Q.-Z., Bao, H., Sanborn, M.E., Irving, A., Ziegler, K., Agee, C., Marti, K., Miao, B., Li, X., Li, Y., Wang, S., 2018. Evidence for a multilayered internal structure of the chondritic acapulcoite-lodranite parent asteroid. Geochim. Cosmochim. Acta 242, 82–101.

Marrocchi, Y., Jones, R.H., Russell, S.S., Hezel, D.C., Barosch, J., Kuznetsova, A., 2024. Chondrule Properties and Formation Conditions. Space Sci. Rev. 220, 69.

Metzler, K., Hezel, D.C., Barosch, J., Wölfer, E., Schneider, J.M., Hellmann, J.L., Berndt, J., Stracke, A., Gattacceca, J., Greenwood, R.C., Franchi, I.A., Burkhardt, C., Kleine, T., 2021. The Loongana (CL) group of carbonaceous chondrites. Geochim. Cosmochim. Acta 304, 1–31.

Millet, M.-A., Dauphas, N., 2014. Ultra-precise titanium stable isotope measurements by double-spike high resolution MC-ICP-MS. J. Anal. At. Spectrom. 29, 1444.

Morbidelli, A., Marrocchi, Y., Ahmad, A.A., Bhandare, A., Charnoz, S., Commercon, B., Dullemond, C.P., Guillot, T., Hennebelle, P., Lee, Y.-N., Lovascio, F., Marschall, R., Marty, B., Maury, A., Tamami, O., 2024. Formation and evolution of a protoplanetary disk: combining observations, simulations and cosmochemical constraints. Astron. Astrophys. 691, A147.

Mougel, B., Moynier, F., Göpel, C., 2018. Chromium isotopic homogeneity between the Moon, the Earth, and enstatite chondrites. Earth Planet. Sci. Lett. 481, 1–8.

Nagashima, K., Krot, A.N., Huss, G.R., 2015. Oxygen-isotope compositions of chondrule phenocrysts and matrix grains in Kakangari K-grouplet chondrite: Implication to a chondrule-matrix genetic relationship. Geochim. Cosmochim. Acta 151, 49–67.

Pape, J., Mezger, K., Bouvier, A.-S., Baumgartner, L.P., 2019. Time and duration of chondrule formation: Constraints from $^{26}$Al-$^{26}$Mg ages of individual chondrules. Geochim. Cosmochim. Acta 244, 416–436.

Pedersen, S.G., Schiller, M., Connelly, J.N., Bizzarro, M., 2019. Testing accretion mechanisms of the H chondrite parent body utilizing nucleosynthetic anomalies. Meteorit. Planet. Sci. 54, 1215–1227.

Petitat, M., Birck, J.-L., Luu, T.H., Gounelle, M., 2011. The chromium isotopic composition of the ungrouped carbonaceous chondrite Tagish Lake. Astrophys. J. 736, 23.

Qin, L., Alexander, C.M.O., Carlson, R.W., Horan, M.F., Yokoyama, T., 2010. Contributors to chromium isotope variation of meteorites. Geochim. Cosmochim. Acta 74, 1122–1145.

Qin, L., Carlson, R.W., Alexander, C.M.O., 2011. Correlated nucleosynthetic isotopic variability in Cr, Sr, Ba, Sm, Nd and Hf in Murchison and QUE 97008. Geochim. Cosmochim. Acta 75, 7806–7828.

Render, J., Brennecka, G.A., Burkhardt, C., Kleine, T., 2022. Solar System evolution and terrestrial planet accretion determined by Zr isotopic signatures of meteorites. Earth Planet. Sci. Lett. 595, 117748.

Render, J., Ebert, S., Burkhardt, C., Kleine, T., Brennecka, G.A., 2019. Titanium isotopic evidence for a shared genetic heritage of refractory inclusions from different carbonaceous chondrites. Geochim. Cosmochim. Acta 254, 40–53.

Render, J., Fischer-Gödde, M., Burkhardt, C., Kleine, T., 2017. The cosmic molybdenum-neodymium isotope correlation and the building material of the Earth. Geochem. Perspect. Lett., 170–178.

Rotaru, M., Birck, J.L., Allègre, C.J., 1992. Clues to early Solar System history from chromium isotopes in carbonaceous chondrites. Nature 358, 465–470.

Rudraswami, N.G., Goswami, J.N., 2007. $^{26}$Al in chondrules from unequilibrated L chondrites: Onset and duration of chondrule formation in the early solar system. Earth Planet. Sci. Lett. 257, 231–244.

Rudraswami, N.G., Goswami, J.N., Chattopadhyay, B., Sengupta, S.K., Thapliyal, A.P., 2008. 26Al records in chondrules from unequilibrated ordinary chondrites: II. Duration of chondrule formation and parent body thermal metamorphism. Earth Planet. Sci. Lett. 274, 93–102.

Rüfenacht, M., Morino, P., Lai, Y.-J., Fehr, M.A., Haba, M.K., Schönbächler, M., 2023. Genetic relationships of solar system bodies based on their nucleosynthetic Ti isotope compositions and sub-structures of the solar protoplanetary disk. Geochim. Cosmochim. Acta 355, 110–125.

Sanborn, M.E., Wimpenny, J., Williams, C.D., Yamakawa, A., Amelin, Y., Irving, A.J., Yin, Q.-Z., 2019. Carbonaceous achondrites Northwest Africa 6704/6693: Milestones for early Solar System chronology and genealogy. Geochim. Cosmochim. Acta 245, 577–596.

Scherer, P., Schultz, L., 2000. Noble gas record, collisional history, and pairing of CV, CO, CK, and other carbonaceous chondrites. Meteorit. Planet. Sci. 35, 145–153.

Schiller, M., Bizzarro, M., Fernandes, V.A., 2018. Isotopic evolution of the protoplanetary disk and the building blocks of Earth and the Moon. Nature 555, 507–510.

Schiller, M., Van Kooten, E., Holst, J.C., Olsen, M.B., Bizzarro, M., 2014. Precise measurement of chromium isotopes by MC-ICPMS. J Anal Spectrom 29, 1406–1416.

Schmitz, B., Yin, Q.-Z., Sanborn, M.E., Tassinari, M., Caplan, C.E., Huss, G.R., 2016. A new type of solar-system material recovered from Ordovician marine limestone. Nat. Commun. 7.

Schneider, J.M., Burkhardt, C., Marrocchi, Y., Brennecka, G.A., Kleine, T., 2020. Early evolution of the solar accretion disk inferred from Cr-Ti-O isotopes in individual chondrules. Earth Planet. Sci. Lett. 551, 116585.

Shirey, S.B., Walker, R.J., 1995. Carius tube digestion for low blank rhenium-osmium analysis. Anal. Chem. 67, 2136–2141.

Shukolyukov, A., Lugmair, G., 2006. Manganese–chromium isotope systematics of carbonaceous chondrites. Earth Planet. Sci. Lett. 250, 200–213.

Shukolyukov, A., Lugmair, G.W., 2004. Manganese-chromium isotope systematics of enstatite meteorites. Geochim. Cosmochim. Acta 68, 2875–2888.

Siron, G., Fukuda, K., Kimura, M., Kita, N.T., 2021. New constraints from 26Al-26Mg chronology of anorthite bearing chondrules in unequilibrated ordinary chondrites. Geochim. Cosmochim. Acta 293, 103–126.

Spitzer, F., Burkhardt, C., Budde, G., Kruijer, T.S., Morbidelli, A., Kleine, T., 2020. Isotopic Evolution of the Inner Solar System Inferred from Molybdenum Isotopes in Meteorites. Astrophys. J. 898, L2.

Spitzer, F., Burkhardt, C., Kruijer, T.S., Kleine, T., 2025a. Comparison of the earliest NC and CC planetesimals: Evidence from ungrouped iron meteorites. Geochim. Cosmochim. Acta 397, 134–148.

Spitzer, F., Hopp, T., Burkhardt, C., Dauphas, N., Kleine, T., 2025b. The evolution of planetesimal reservoirs revealed by Fe-Ni isotope anomalies in differentiated meteorites. Earth Planet. Sci. Lett. 667, 119530.

Steele, R.C.J., Boehnke, P., 2015. Titanium isotope source relations and the extent of mixing in the proto-solar nebula examined by independent component analysis. Astrophys. J. 802, 80.

Stracke, A., Palme, H., Gellissen, M., Münker, C., Kleine, T., Birbaum, K., Günther, D., Bourdon, B., Zipfel, J., 2012. Refractory element fractionation in the Allende meteorite: Implications for solar nebula condensation and the chondritic composition of planetary bodies. Geochim. Cosmochim. Acta 85, 114–141.

Sugiura, N., Fujiya, W., 2014. Correlated accretion ages and $\varepsilon^{54}$Cr of meteorite parent bodies and the evolution of the solar nebula. Meteorit. Planet. Sci. 49, 772–787.

Tissot, F.L.H., Burkhardt, C., Kuznetsova, A., Pack, A., Schiller, M., Spitzer, F., Van Kooten, E.M.M.E., Yap, T.E., 2025. Infall and Disk Processes – the Message from Meteorites. Space Sci. Rev. 221, 85.

Torrano, Z.A., Alexander, C.M.O., Carlson, R.W., Render, J., Brennecka, G.A., Bullock, E.S., 2024. A common isotopic reservoir for amoeboid olivine aggregates (AOAs) and calcium-aluminum-rich inclusions (CAIs) revealed by Ti and Cr isotopic compositions. Earth Planet. Sci. Lett. 627, 118551.

Torrano, Z.A., Brennecka, G.A., Mercer, C.M., Romaniello, S.J., Rai, V.K., Hines, R.R., Wadhwa, M., 2023. Titanium and chromium isotopic compositions of calcium-aluminum-rich inclusions: Implications for the sources of isotopic anomalies and the formation of distinct isotopic reservoirs in the early Solar System. Geochim. Cosmochim. Acta 348, 309–322.

Torrano, Z.A., Brennecka, G.A., Williams, C.D., Romaniello, S.J., Rai, V.K., Hines, R.R., Wadhwa, M., 2019. Titanium isotope signatures of calcium-aluminum-rich inclusions from CV and CK chondrites: Implications for early Solar System reservoirs and mixing. Geochim. Cosmochim. Acta 263, 13–30.

Torrano, Z.A., Schrader, D.L., Davidson, J., Greenwood, R.C., Dunlap, D.R., Wadhwa, M., 2021. The relationship between CM and CO chondrites: Insights from combined analyses of titanium, chromium, and oxygen isotopes in CM, CO, and ungrouped chondrites. Geochim. Cosmochim. Acta 301, 70–90.

Touboul, M., Sprung, P., Aciego, S.M., Bourdon, B., Kleine, T., 2015. Hf–W chronology of the eucrite parent body. Geochim. Cosmochim. Acta 156, 106–121.

Trieloff, M., Jessberger, E.K., Herrwerth, I., Hopp, J., Fiéni, C., Ghélis, M., Bourot-Denise, M., Pellas, P., 2003. Structure and thermal history of the H-chondrite parent asteroid revealed by thermochronometry. Nature 422, 502–506.

Trinquier, A., Birck, J., Allegre, C.J., 2007. Widespread $^{54}$Cr Heterogeneity in the Inner Solar System. Astrophys. J. 655, 1179–1185.

Trinquier, A., Birck, J.-L., Allègre, C.J., 2008a. High-precision analysis of chromium isotopes in terrestrial and meteorite samples by thermal ionization mass spectrometry. J. Anal. At. Spectrom. 23, 1565.

Trinquier, A., Birck, J.-L., Allègre, C.J., Göpel, C., Ulfbeck, D., 2008b. $^{53}$Mn–$^{53}$Cr systematics of the early Solar System revisited. Geochim. Cosmochim. Acta 72, 5146–5163.

Trinquier, A., Elliott, T., Ulfbeck, D., Coath, C., Krot, A.N., Bizzarro, M., 2009. Origin of Nucleosynthetic Isotope Heterogeneity in the Solar Protoplanetary Disk. Science 324, 374–376.

Van Kooten, E., Cavalcante, L., Wielandt, D., Bizzarro, M., 2020. The role of Bells in the continuous accretion between the CM and CR chondrite reservoirs. Meteorit. Planet. Sci. 55, 575–590.

Van Kooten, E.M.M.E., Wielandt, D., Schiller, M., Nagashima, K., Thomen, A., Larsen, K.K., Olsen, M.B., Nordlund, Å., Krot, A.N., Bizzarro, M., 2016. Isotopic evidence for primordial molecular cloud material in metal-rich carbonaceous chondrites. Proc. Natl. Acad. Sci. 113, 2011–2016.

Villeneuve, J., Chaussidon, M., Libourel, G., 2009. Homogeneous Distribution of $^{26}$Al in the Solar System from the Mg Isotopic Composition of Chondrules. Science 325, 985–988.

Wang, K., Moynier, F., Podosek, F., Foriel, J., 2011. $^{58}$Fe and $^{54}$Cr in early solar system materials. Astrophys. J. 739, L58.

Warren, P.H., 2011a. Stable-isotopic anomalies and the accretionary assemblage of the Earth and Mars: A subordinate role for carbonaceous chondrites. Earth Planet. Sci. Lett. 311, 93–100.

Warren, P.H., 2011b. Stable isotopes and the noncarbonaceous derivation of ureilites, in common with nearly all differentiated planetary materials. Geochim. Cosmochim. Acta 75, 6912–6926.

Warren, P.H., Rubin, A.E., Isa, J., Brittenham, S., Ahn, I., Choi, B.-G., 2013. Northwest Africa 6693: A new type of FeO-rich, low-$\Delta^{17}$O, poikilitic cumulate achondrite. Geochim. Cosmochim. Acta 107, 135–154.

Weisberg, M.K., Ebel, D.S., Nakashima, D., Kita, N.T., Humayun, M., 2015. Petrology and geochemistry of chondrules and metal in NWA 5492 and GRO 95551: A new type of metal-rich chondrite. Geochim. Cosmochim. Acta 167, 269–285.

Weisberg, M.K., McCoy, T.J., Krot, A.N., 2006. Systematics and Evaluation of Meteorite Classification. In Meteorites and the Early Solar System II Univ_of_Arizona_Press. p. 34.

Weisberg, M.K., Prinz, M., Clayton, R.N., Mayeda, T.K., Grady, M.M., Franchi, I., Pillinger, C.T., Kallemeyn, G.W., 1996. The K (Kakangari) chondrite grouplet. Geochim. Cosmochim. Acta 60, 4253–4263.

Williams, C.D., Sanborn, M.E., Defouilloy, C., Yin, Q.-Z., Kita, N.T., Ebel, D.S., Yamakawa, A., Yamashita, K., 2020. Chondrules reveal large-scale outward transport of inner Solar System materials in the protoplanetary disk. Proc. Natl. Acad. Sci. 117, 23426–23435.

Williams, N.H., Fehr, M.A., Parkinson, I.J., Mandl, M.B., Schönbächler, M., 2021. Titanium isotope fractionation in solar system materials. Chem. Geol. 568, 120009.

Wölfer, E., Budde, G., Kleine, T., 2023. Age and genetic relationships among CB, CH and CR chondrites. Geochim. Cosmochim. Acta, S0016703723004842.

Worsham, E.A., Bermingham, K.R., Walker, R.J., 2017. Characterizing cosmochemical materials with genetic affinities to the Earth: Genetic and chronological diversity within the IAB iron meteorite complex. Earth Planet. Sci. Lett. 467, 157–166.

Xu, L.-J., Zhu, K., Man, Q.-R., Lewis, J., Ma, H., Liu, S.-A., 2023. Precise and Accurate Mass-independent Chromium Isotope Measurement by Total Evaporation Mode on Thermal Ionization Mass Spectrometry (TE-TIMS) at 200 ng Level. At. Spectrosc. 44(3), 142–152.

Yamakawa, A., Yamashita, K., Makishima, A., Nakamura, E., 2010. Chromium isotope systematics of achondrites: chronology and isotopic heterogeneity of the inner Solar System bodies. Astrophys. J. 720, 150–154.

Yamashita, K., Maruyama, S., Yamakawa, A., Nakamura, E., 2010. $^{53}$Mn-$^{53}$Cr Chronometry of CB chondrite: Evidence for uniform distribution of $^{53}$Mn in the early Solar System. Astrophys. J. 723, 20–24.

Yang, B., Fang, Z., Liu, J., Heck, P.R., Qin, L., 2025a. Iron meteorites reveal differential material mixing patterns in the inner and outer protoplanetary disk. Earth Planet. Sci. Lett. 671, 119664.

Yang, B., Pang, R., Wang, Q., Zhang, A., Du, W., Qin, L., 2025b. Highly Heterogeneous Parent Body of the Rare Andesitic Erg Chech 002 Meteorite Revealed by the Revisited Mn–Cr Isotopic Systematics. Planet. Sci. J. 6, 73.

Yokoyama, T., Nagai, Y., Fukai, R., Hirata, T., 2019. Origin and Evolution of Distinct Molybdenum Isotopic Variabilities within Carbonaceous and Noncarbonaceous Reservoirs. Astrophys. J. 883, 62.

Yokoyama, T., Wadhwa, M., Iizuka, T., Rai, V., Gautam, I., Hibiya, Y., Masuda, Y., Haba, M.K., Fukai, R., Hines, R., Phelan, N., Abe, Y., Aléon, J., Alexander, C.M.O., Amari, S., Amelin, Y., Bajo, K., Bizzarro, M., Bouvier, A., Carlson, R.W., Chaussidon, M., Choi, B.-G., Dauphas, N., Davis, A.M., Di Rocco, T., Fujiya, W., Hidaka, H., Homma, H., Hoppe, P., Huss, G.R., Ichida, K., Ireland, T., Ishikawa, A., Itoh, S., Kawasaki, N., Kita, N.T., Kitajima, K., Kleine, T., Komatani, S., Krot, A.N., Liu, M.-C., McKeegan, K.D., Morita, M., Motomura, K., Moynier, F., Nakai, I., Nagashima, K., Nguyen, A., Nittler, L., Onose, M., Pack, A., Park, C., Piani, L., Qin, L., Russell, S., Sakamoto, N., Schönbächler, M., Tafla, L., Tang, H., Terada, K., Terada, Y., Usui, T., Wada, S., Walker, R.J., Yamashita, K., Yin, Q.-Z., Yoneda, S., Young, E.D., Yui, H., Zhang, A.-C., Nakamura, T., Naraoka, H., Noguchi, T., Okazaki, R., Sakamoto, K., Yabuta, H., Abe, M., Miyazaki, A., Nakato, A., Nishimura, M., Okada, T., Yada, T., Yogata, K., Nakazawa, S., Saiki, T., Tanaka, S., Terui, F., Tsuda, Y., Watanabe, S., Yoshikawa, M., Tachibana, S., Yurimoto, H., 2023. Water circulation in Ryugu asteroid affected the distribution of nucleosynthetic isotope anomalies in returned sample. Sci. Adv. 9, eadi7048.

Zhang, J., Dauphas, N., Davis, A.M., Leya, I., Fedkin, A., 2012. The proto-Earth as a significant source of lunar material. Nat. Geosci. 5, 251–255.

Zhang, J., Dauphas, N., Davis, A.M., Pourmand, A., 2011. A new method for MC-ICPMS measurement of titanium isotopic composition: Identification of correlated isotope anomalies in meteorites. J. Anal. At. Spectrom. 26, 2197.

Zhu, K., Moynier, F., Schiller, M., Alexander, C.M.O., Barrat, J.-A., Bischoff, A., Bizzarro, M., 2021a. Mass-independent and mass-dependent Cr isotopic composition of the Rumuruti (R) chondrites: Implications for their origin and planet formation. Geochim. Cosmochim. Acta 293, 598–609.

Zhu, K., Moynier, F., Schiller, M., Alexander, C.M.O., Davidson, J., Schrader, D.L., van Kooten, E., Bizzarro, M., 2021b. Chromium isotopic insights into the origin of chondrite parent bodies and the early terrestrial volatile depletion. Geochim. Cosmochim. Acta 301, 158–186.

Zhu, K., Moynier, F., Schiller, M., Becker, H., Barrat, J.-A., Bizzarro, M., 2021c. Tracing the origin and core formation of the enstatite achondrite parent bodies using Cr isotopes. Geochim. Cosmochim. Acta 308, 256–272.

Zhu, K., Moynier, F., Schiller, M., Wielandt, D., Larsen, K.K., van Kooten, E.M.M.E., Barrat, J.-A., Bizzarro, M., 2020. Chromium Isotopic Constraints on the Origin of the Ureilite Parent Body. Astrophys. J. 888, 126.

Zhu, K., Moynier, F., Wielandt, D., Larsen, K.K., Barrat, J.-A., Bizzarro, M., 2019. Timing and Origin of the Angrite Parent Body Inferred from Cr Isotopes. Astrophys. J. 877, L13.

Zhu, K., Schiller, M., Moynier, F., Groen, M., Alexander, C.M.O., Davidson, J., Schrader, D.L., Bischoff, A., Bizzarro, M., 2023. Chondrite diversity revealed by chromium, calcium and magnesium isotopes. Geochim. Cosmochim. Acta 342, 156–168.

Zhu, K., Schiller, M., Pan, L., Saji, N.S., Larsen, K.K., Amsellem, E., Rundhaug, C., Sossi, P., Leya, I., Moynier, F., Bizzarro, M., 2022. Late delivery of exotic chromium to the crust of Mars by water-rich carbonaceous asteroids. Sci. Adv. 8, eabp8415.

Zhu, K., Nakanishi, N., Render, J., Shollenberger, Q.R., Yokoyama, T., Ishikawa, A., Chen, L., 2025. “CY1” Chondrites Produced by Impact Dehydration of the CI Chondrite Parent Body. ApJL 984, L54.

Zhu, K., Schiller, M., Moynier, F., Groen, M., M. O’D. Alexander, C., Davidson, J., Schrader, D.L., Bischoff, A., Bizzarro, M., 2022. Chondrite diversity revealed by chromium, calcium and magnesium isotopes. Geochim. Cosmochim. Acta, S0016703722006615.

**Table 1:** Titanium, Cr, and Mo isotope data for non-carbonaceous meteorites and terrestrial samples analyzed in this study.

| Group | Sample name | Wt. (mg) | s.aq. | p.aq. | Ti (µg/g) | N | ε46Ti (± 2σ) | ε48Ti (± 2σ) | ε50Ti (± 2σ) | Cr (µg/g) | N | ε53Cr (± 2σ) | ε54Cr (± 2σ) | Mo (µg/g) | N | ε94Mo (± 2σ) | ε95Mo (± 2σ) | Δ95Mo (± 2σ) | Ref. |
|---|---|---|---|---|---|---|---|---|---|---|---|---|---|---|---|---|---|---|---|
| ***Terrestrial rocks*** | | | | | | | | | | | | | | | | | | | |
| Ocean island basalt | BHVO-2 | | | | 16368 | 61 | –0.01 ± 0.04 | 0.02 ± 0.03 | –0.10 ± 0.03 | 287 | 5 | 0.14 ± 0.09 | 0.20 ± 0.12 | 4.07 | 40 | 0.03 ± 0.04 | 0.04 ± 0.02 | 2 ± 3 | [1] |
| Arc basalt | JB-2 | | | | 6994 | 127 | –0.04 ± 0.02 | 0.01 ± 0.02 | –0.02 ± 0.03 | | | | | 1.01 | 15 | 0.06 ± 0.07 | 0.13 ± 0.04 | 9 ± 6 | [1] |
| Andesite | JA-2 | | | | 4013 | 13 | –0.12 ± 0.07 | 0.01 ± 0.05 | 0.01 ± 0.09 | 425 | 9 | 0.04 ± 0.05 | 0.17 ± 0.12 | 0.58 | 17 | 0.09 ± 0.04 | 0.12 ± 0.04 | 7 ± 5 | [1] |
| Dunite | DTS-2a | | | | | | | | | 16003 | 6 | 0.08 ± 0.04 | 0.13 ± 0.16 | | | | | | |
| | DTS-2b | | | | | | | | | 15500 | 34 | 0.08 ± 0.04 | 0.07 ± 0.06 | 0.06 | 4 | 0.02 ± 0.24 | 0.10 ± 0.12 | 9 ± 19 | [1] |
| | replicate | | | | | | | | | 15500 | 8 | 0.05 ± 0.02 | 0.08 ± 0.13 | 0.06 | 4 | 0.02 ± 0.24 | 0.10 ± 0.12 | 9 ± 19 | [1] |
| ***NC bulk meteorites*** | | | | | | | | | | | | | | | | | | | |
| Rumuruti chondrites | Rumuruti | 30 | | | 573 | 12 | –0.07 ± 0.08 | –0.06 ± 0.03 | –0.39 ± 0.07 | | | | | | | | | | |
| | NWA 753 | s. aq. | [1] | | 350 | 12 | –0.10 ± 0.05 | 0.02 ± 0.06 | –0.38 ± 0.08 | 3543 | 8 | 0.25 ± 0.05 | –0.07 ± 0.09 | 0.57 | 3 | 0.65 ± 0.22 | 0.31 ± 0.15 | –8 ± 20 | [1] |
| | NWA 6145 | 32 | | [1] | 549 | 13 | –0.10 ± 0.07 | 0.00 ± 0.05 | –0.45 ± 0.13 | 3866 | 10 | 0.24 ± 0.05 | –0.11 ± 0.14 | 0.72 | 3 | 0.49 ± 0.22 | 0.23 ± 0.15 | –6 ± 20 | [1] |
| | NWA 6145 | 33 | | [1] | 511 | 13 | –0.10 ± 0.07 | –0.09 ± 0.03 | –0.44 ± 0.14 | 3437 | 10 | 0.18 ± 0.06 | –0.13 ± 0.14 | 0.72 | 3 | 0.49 ± 0.22 | 0.23 ± 0.15 | –6 ± 20 | [1] |
| Kakangari chondrites | LEW 87232 | 526 | | | 408 | 16 | –0.26 ± 0.06 | –0.06 ± 0.06 | –1.28 ± 0.08 | 1052 | 9 | 0.61 ± 0.04 | –0.41 ± 0.09 | 1.41 | 6 | 0.94 ± 0.12 | 0.37 ± 0.09 | –19 ± 11 | |
| Ungrouped chondrites | NWA 13202 | 25/168 | | | 615 | 10 | –0.08 ± 0.04 | –0.01 ± 0.03 | –0.64 ± 0.05 | 3084 | 2 | 0.28 ± 0.19 | –0.24 ± 0.17 | 2.39 | 4 | 0.69 ± 0.16 | 0.39 ± 0.16 | –2 ± 19 | |
| | NWA 13202† | | | | | | | | | | | | | | | 0.73 ± 0.16 | 0.45 ± 0.16 | 1 ± 19 | |
| | NWA 5492 | 227 | | | | | | | | | | | | 2.78 | 6 | 0.33 ± 0.16 | 0.18 ± 0.06 | –2 ± 11 | |
| Acapulcoites | Dho 125 | s. aq. | [1] | | 629 | 13 | –0.28 ± 0.09 | 0.00 ± 0.03 | –1.40 ± 0.06 | 3450 | 12 | 0.29 ± 0.04 | –0.49 ± 0.12 | 0.91 | 5 | 0.94 ± 0.12 | 0.41 ± 0.07 | –15 ± 10 | [1] |
| | NWA 11048 | s. aq. | [2] | | 420 | 12 | –0.30 ± 0.09 | –0.03 ± 0.05 | –1.42 ± 0.16 | | | | | 1.34 | 5 | 0.56 ± 0.11 | 0.28 ± 0.12 | –5 ± 14 | [2] |
| | MET 01195 | s. aq. | [2] | | 515 | 13 | –0.23 ± 0.06 | 0.04 ± 0.03 | –1.39 ± 0.09 | 4654 | 8 | 0.28 ± 0.10 | –0.62 ± 0.06 | 1.14 | 5 | 0.89 ± 0.09 | 0.49 ± 0.03 | –4 ± 6 | [2] |
| Lodranites | GRA 95209 | 70 | | [3] | 467 | 13 | –0.31 ± 0.05 | 0.00 ± 0.02 | –1.44 ± 0.07 | 2508 | 8 | 0.37 ± 0.05 | –0.57 ± 0.11 | | 1 | 1.10 ± 0.30 | 0.48 ± 0.15 | –18 ± 23 | [3] |
| | NWA 7474 | 82 | | | 361 | 13 | –0.33 ± 0.05 | 0.03 ± 0.04 | –1.34 ± 0.06 | 1931 | 10 | 0.40 ± 0.09 | –0.23 ± 0.12 | | | | | | |
| | NWA 7474† | | | | | | | | | | | 0.32 ± 0.03 | –0.55 ± 0.12 | | | | | | |
| Brachinites | NWA 3151 | s. aq. | [1] | | 58 | 13 | –0.29 ± 0.09 | –0.15 ± 0.03 | –1.62 ± 0.08 | 4232 | 8 | 0.19 ± 0.08 | –0.41 ± 0.09 | 0.47 | 3 | 1.14 ± 0.22 | 0.61 ± 0.15 | –7 ± 20 | [1] |
| | NWA 4882 | s. aq. | [1] | | 63 | 11 | –0.27 ± 0.08 | 0.00 ± 0.05 | –1.27 ± 0.08 | | | | | 0.38 | 4 | 1.10 ± 0.21 | 0.56 ± 0.11 | –10 ± 17 | [1] |
| | NWA 10637 | s. aq. | [2] | | 136 | 11 | –0.22 ± 0.07 | –0.17 ± 0.08 | –1.36 ± 0.10 | 3275 | 4 | 0.30 ± 0.07 | –0.30 ± 0.01 | 1.05 | 7 | 1.22 ± 0.09 | 0.58 ± 0.07 | –15 ± 9 | [2] |
| Winonaites | HaH 193 | 41 | | [2] | 635 | 13 | –0.05 ± 0.05 | 0.05 ± 0.05 | –0.41 ± 0.13 | 853 | 11 | 0.55 ± 0.05 | 0.51 ± 0.14 | 0.43 | 3 | 0.29 ± 0.22 | 0.14 ± 0.15 | –3 ± 20 | [2] |
| | HaH 193† | | | | | | | | | | | 0.42 ± 0.05 | –0.02 ± 0.14 | | | | | | |
| | replicate | 80 | | [2] | | | | | | 856 | 9 | 0.55 ± 0.04 | 0.49 ± 0.09 | 0.43 | 3 | 0.29 ± 0.22 | 0.14 ± 0.15 | –3 ± 20 | [2] |
| | replicate† | | | | | | | | | | | 0.42 ± 0.05 | –0.03 ± 0.09 | | | | | | |
| Ureilites | NWA 5938 | 561 | | | 98 | 14 | –0.33 ± 0.06 | 0.02 ± 0.11 | –1.87 ± 0.14 | 5105 | 9 | 0.24 ± 0.05 | –0.70 ± 0.13 | | | | | | |
| | replicate | 938 | | | 106 | 16 | –0.41 ± 0.06 | –0.01 ± 0.07 | –2.19 ± 0.10 | | | | | | | | | | |
| | DaG 999 | 508 | | | 177 | 14 | –0.35 ± 0.07 | –0.01 ± 0.09 | –1.88 ± 0.09 | | | | | | | | | | |
| | Sahara 98501 | 539 | | | 161 | 14 | –0.35 ± 0.05 | 0.08 ± 0.11 | –1.98 ± 0.08 | | | | | | | | | | |
| | Dho 132 | 446 | | [1] | 225 | 14 | –0.34 ± 0.06 | 0.09 ± 0.08 | –2.07 ± 0.08 | 4792 | | | | 0.34 | 3 | 0.65 ± 0.22 | 0.35 ± 0.15 | –4 ± 20 | [1] |
| | HaH 064 | 513 | | | 203 | 14 | –0.34 ± 0.09 | 0.07 ± 0.09 | –2.00 ± 0.11 | | | | | | | | | | |
| | MS-MU-16 | 940 | | | 334 | 12 | –0.43 ± 0.12 | –0.04 ± 0.05 | –2.02 ± 0.08 | 4430 | 12 | 0.25 ± 0.06 | –0.71 ± 0.12 | | | | | | |
| | MS-MU-17 | s. aq. | [1] | | 458 | 12 | –0.35 ± 0.06 | 0.00 ± 0.03 | –1.99 ± 0.11 | | | | | 0.13 | 1 | 0.93 ± 0.22 | 0.42 ± 0.15 | –13 ± 20 | [1] |
| | MS-MU-20 | s. aq. | [1] | | 68 | 12 | –0.44 ± 0.07 | –0.05 ± 0.03 | –2.10 ± 0.08 | 4820 | 10 | 0.20 ± 0.04 | –0.74 ± 0.08 | 0.32 | 2 | 1.15 ± 0.22 | 0.51 ± 0.15 | –18 ± 20 | [1] |
| | ALM-A | 525 | | | 3864 | 12 | –0.44 ± 0.10 | 0.03 ± 0.03 | –1.94 ± 0.07 | | | | | | | | | | |
| Prim. enstatite achond. | NWA 2526 | s. aq. | | [2] | 75 | 12 | –0.03 ± 0.05 | 0.07 ± 0.06 | –0.01 ± 0.13 | 700 | 9 | 0.17 ± 0.05 | 0.08 ± 0.11 | 0.6 | 4 | 0.60 ± 0.13 | 0.39 ± 0.13 | 3 ± 15 | [2] |
| Ungrouped achond. | NWA 1058 | s. aq. | [1] | | 406 | 12 | –0.20 ± 0.04 | –0.06 ± 0.05 | –0.68 ± 0.10 | 1589 | 10 | 0.50 ± 0.07 | –0.39 ± 0.06 | 1.41 | 6 | 1.31 ± 0.11 | 0.68 ± 0.09 | –10 ± 11 | [1] |
| | GRA 06128 | 549 | | | 411 | 19 | –0.20 ± 0.06 | –0.26 ± 0.13 | –1.71 ± 0.06 | 146 | 11 | 1.24 ± 0.05 | –0.63 ± 0.14 | 0.1 | 1 | 1.03 ± 0.22 | 0.60 ± 0.15 | –1 ± 20 | |
| | NWA 6112 | s. aq. | [2] | | 304 | 12 | –0.27 ± 0.10 | –0.07 ± 0.04 | –1.44 ± 0.09 | 7132 | 14 | 0.24 ± 0.04 | –0.60 ± 0.13 | 0.89 | 3 | 1.55 ± 0.22 | 0.79 ± 0.15 | –13 ± 20 | [2] |
| | NWA 5400 | s. aq. | [2] | | 186 | 12 | –0.19 ± 0.09 | –0.08 ± 0.04 | –1.00 ± 0.07 | 5258 | 8 | 0.18 ± 0.04 | –0.46 ± 0.07 | 0.57 | 3 | 0.66 ± 0.22 | 0.31 ± 0.15 | –8 ± 20 | [2] |
| Angrites | NWA 4931 | s. aq. | [1] | | 4195 | 11 | –0.20 ± 0.09 | –0.04 ± 0.04 | –1.30 ± 0.08 | | | | | 0.31 | 5 | 0.75 ± 0.11 | 0.39 ± 0.06 | –6 ± 9 | [1] |
| Aubrites | Peña Blanca Spring | s. aq. | [1] | | 322 | 14 | –0.12 ± 0.05 | 0.03 ± 0.05 | –0.05 ± 0.09 | 394 | | | | 0.01 | 1 | 0.58 ± 0.22 | 0.16 ± 0.15 | –19 ± 20 | [1] |

| | | | | | | | | | | | | | | | | | | |
|---|---|---|---|---|---|---|---|---|---|---|---|---|---|---|---|---|---|---|
| | replicate | s. aq. | [1] | 301 | 11 | –0.12 ± 0.12 | 0.04 ± 0.03 | –0.09 ± 0.08 | 394 | | | | 0.01 | 1 | 0.58 ± 0.22 | 0.16 ± 0.15 | –19 ± 20 | [1] |
| Eucrites | NWA 11050 | 1053 | | 1199 | 13 | –0.27 ± 0.05 | 0.05 ± 0.02 | –1.22 ± 0.11 | | | | | | | | | | |
| Diogenites | NWA 11049 | 903 | | 528 | 12 | –0.22 ± 0.07 | 0.03 ± 0.05 | –1.25 ± 0.10 | | | | | | | | | | |
| Mesosiderites | Acfer 063 | s. aq. | [1] | 222 | 12 | –0.24 ± 0.10 | –0.04 ± 0.08 | –1.24 ± 0.10 | 2864 | 10 | 0.34 ± 0.03 | –0.55 ± 0.11 | 2.37 | 6 | 1.05 ± 0.12 | 0.46 ± 0.07 | –17 ± 10 | [1] |
| | Ilafegh 002 | s. aq. | [1] | 257 | 12 | –0.29 ± 0.10 | –0.03 ± 0.05 | –1.19 ± 0.06 | | | | | 2.43 | 7 | 1.01 ± 0.16 | 0.45 ± 0.12 | –15 ± 15 | [1] |
| | NWA 2538 | s. aq. | [1] | 162 | 12 | –0.33 ± 0.05 | –0.07 ± 0.02 | –1.34 ± 0.07 | | | | | 1.81 | 5 | 1.04 ± 0.17 | 0.49 ± 0.12 | –13 ± 16 | [1] |

Titanium isotope ratios are internally normalized to $^{49}$Ti/$^{47}$Ti = 0.749766 and reported relative to the OL-Ti bracketing standard; Cr isotope ratios are internally normalized to $^{50}$Cr/$^{52}$Cr = 0.051859 and reported relative to the NIST SRM3112a standard; Mo isotope ratios are internally normalized to $^{98}$Mo/$^{96}$Mo = 1.453173 and reported relative to the Alfa Aesar Mo standard. $\Delta^{95}$Mo = ($\varepsilon^{95}$Mo – 0.596 × $\varepsilon^{94}$Mo) × 100; The $\Delta^{95}$Mo notation reflects the ppm-deviation of a sample's $^{94}$Mo and $^{95}$Mo isotope composition from an *s*-process mixing line passing through the origin in $\varepsilon^{95}$Mo vs. $\varepsilon^{94}$Mo space. Given uncertainties of individual samples are Student-t 95% confidence intervals (95% CI) for samples with N>3 [i.e., ($t_{0.95,\,N-1}$ × s.d.)/√N] or reflect the external reproducibility (2 s.d.) obtained from repeated analysis of the terrestrial standards for samples with N≤3 (in case of Mo). N: number of analyses; s. aq./p. aq.: solution aliquot (equivalent to ~30 µg Ti or Cr) or powder aliquot taken from already digested/grinded samples that have previously been analyzed for Mo.

† Meteorite sample corrected for cosmic ray exposure effects.

Ti, Cr, and Mo concentrations as determined by quadrupole ICP–MS, which have an uncertainty of ~10%.

References: [1] Budde et al., 2019, [2] Hopp et al., 2020, [3] Worsham et al., 2017.

**Table 2:** Titanium, Cr, and Mo isotope data for carbonaceous meteorites analyzed in this study.

| Group | Sample name | Wt. (mg) | Ti (µg/g) | N | $\varepsilon^{46}$Ti (± 2σ) | $\varepsilon^{48}$Ti (± 2σ) | $\varepsilon^{50}$Ti (± 2σ) | N | $\varepsilon^{53}$Cr (± 2σ) | $\varepsilon^{54}$Cr (± 2σ) | Mo (µg/g) | N | $\varepsilon^{94}$Mo (± 2σ) | $\varepsilon^{95}$Mo (± 2σ) | $\Delta^{95}$Mo (± 2σ) |
|---|---|---|---|---|---|---|---|---|---|---|---|---|---|---|---|
| ***CC bulk meteorites*** | | | | | | | | | | | | | | | |
| Bulk Allende | Allende (MS-A) | 32 | 895 | 12 | 0.70 ± 0.06 | -0.06 ± 0.04 | 3.21 ± 0.11 | | | | | | | | |
| | Allende (MS-A) | 24 | 891 | 11 | 0.63 ± 0.09 | -0.02 ± 0.05 | 3.22 ± 0.08 | | | | | | | | |
| | Allende (MS-A) | 31 | 901 | 16 | 0.50 ± 0.07 | -0.05 ± 0.07 | 3.39 ± 0.09 | | | | | | | | |
| Allende subsamples | A2 | 15 | 1019 | 13 | 0.64 ± 0.04 | 0.06 ± 0.05 | 3.80 ± 0.07 | | | | | | | | |
| | A4 | 30 | 779 | 13 | 0.47 ± 0.10 | -0.01 ± 0.02 | 2.22 ± 0.11 | | | | | | | | |
| | A6 | 26 | 809 | 11 | 0.36 ± 0.06 | -0.04 ± 0.02 | 2.35 ± 0.06 | | | | | | | | |
| | B1 | 22 | 755 | 13 | 0.44 ± 0.07 | -0.02 ± 0.04 | 2.84 ± 0.08 | | | | | | | | |
| | B6 | 11 | 803 | 11 | 0.41 ± 0.06 | 0.04 ± 0.04 | 2.90 ± 0.06 | | | | | | | | |
| | C5 | 23 | 1115 | 13 | 0.71 ± 0.08 | -0.01 ± 0.06 | 4.03 ± 0.07 | | | | | | | | |
| | D1 | 19 | 803 | 11 | 0.45 ± 0.06 | -0.12 ± 0.04 | 3.06 ± 0.06 | | | | | | | | |
| | D4 | 22 | 809 | 14 | 0.41 ± 0.07 | 0.00 ± 0.04 | 2.16 ± 0.10 | | | | | | | | |
| | E2 | 19 | 899 | 13 | 0.50 ± 0.05 | 0.01 ± 0.04 | 2.98 ± 0.09 | | | | | | | | |
| | F2 | 26 | 827 | 14 | 0.46 ± 0.07 | -0.06 ± 0.04 | 2.91 ± 0.10 | | | | | | | | |
| | G1 | 23 | 989 | 13 | 0.59 ± 0.07 | -0.06 ± 0.09 | 3.01 ± 0.06 | | | | | | | | |
| | G2 | 25 | 887 | 11 | 0.61 ± 0.07 | 0.06 ± 0.04 | 3.78 ± 0.08 | | | | | | | | |
| CM | Moapa Valley | 43 | 631 | 14 | 0.48 ± 0.09 | 0.02 ± 0.05 | 2.94 ± 0.08 | | | | | | | | |
| CL | NWA 13400 | 916 | 900 | 12 | 0.53 ± 0.08 | -0.05 ± 0.04 | 2.73 ± 0.11 | | | | 1.32 | 9 | 1.14 ± 0.10 | 0.99 ± 0.06 | 31 ± 8 |
| Ungrouped chondrites | DaG 430 | 25 | 827 | 15 | 0.48 ± 0.06 | -0.01 ± 0.04 | 2.23 ± 0.12 | | | | | | | | |
| | DaG 055 | 49 | 761 | 5 | | | 2.84 ± 0.17 | | | | | | | | |
| | NWA 11086 | 31 | 563 | 15 | 0.48 ± 0.07 | -0.07 ± 0.04 | 2.68 ± 0.11 | | | | | | | | |
| | NWA 12957 | 14 | 519 | 14 | 0.42 ± 0.08 | 0.01 ± 0.02 | 3.05 ± 0.10 | | | | | | | | |
| | NWA 12416 | 13 | 620 | 14 | 0.44 ± 0.09 | 0.00 ± 0.05 | 3.08 ± 0.07 | | | | | | | | |
| | NWA 5958 | 15 | 563 | 15 | 0.45 ± 0.06 | 0.01 ± 0.04 | 2.81 ± 0.11 | | | | | | | | |
| | Acfer 094 | 14 | 592 | 15 | 0.38 ± 0.07 | 0.05 ± 0.05 | 3.41 ± 0.09 | | | | | | | | |
| | GRO 95577 | 13 | 522 | 15 | 0.32 ± 0.08 | -0.06 ± 0.03 | 1.91 ± 0.09 | | | | | | | | |
| | NWA 11024 | 33 | 661 | 15 | 0.45 ± 0.07 | 0.00 ± 0.04 | 2.81 ± 0.10 | | | | | | | | |
| | Essebi | 32 | 565 | 15 | 0.45 ± 0.04 | 0.02 ± 0.03 | 2.97 ± 0.09 | | | | | | | | |
| Ungrouped achondrites | NWA 6926† | s. aq. | 226 | 12 | 0.41 ± 0.09 | -0.09 ± 0.05 | 1.92 ± 0.13 | 9 | 0.26 ± 0.05 | 1.44 ± 0.11 | 0.91 | 8 | 1.48 ± 0.12 | 1.14 ± 0.07 | 26 ± 10 |

Titanium isotope ratios are internally normalized to $^{49}$Ti/$^{47}$Ti = 0.749766 and reported relative to the OL-Ti bracketing standard; Cr isotope ratios are internally normalized to $^{50}$Cr/$^{52}$Cr = 0.051859 and reported relative to the NIST SRM3112a standard; Mo isotope ratios are internally normalized to $^{98}$Mo/$^{96}$Mo = 1.453173 and reported relative to the Alfa Aesar Mo standard. $\Delta^{95}$Mo = ($\varepsilon^{95}$Mo – 0.596 × $\varepsilon^{94}$Mo) × 100; The $\Delta^{95}$Mo notation reflects the ppm-deviation of a sample's $^{94}$Mo and $^{95}$Mo isotope composition from an *s*-process mixing line passing through the origin in $\varepsilon^{95}$Mo vs. $\varepsilon^{94}$Mo space. Given uncertainties of individual samples are Student-t 95% confidence intervals (95% CI; i.e., ($t_{0.95,\,N-1}$ × s.d.)/$\sqrt{N}$). N: number of analyses. Titanium and Mo concentrations as determined by quadrupole ICP–MS, which have an uncertainty of ~10%. † For NWA 6926, a solution aliquot equivalent to ~30 µg Ti was taken from the digestion of Hopp et al. (2020), who reported the Mo isotopic data of this sample.

**Table 3:** Titanium and Mo isotope data for the ordinary chondrite leachate fractions analyzed in this study.

| Sample | Fraction | Ti (µg/g) | N | ε46Ti (± 2σ) | ε48Ti (± 2σ) | ε50Ti (± 2σ) | Mo (µg/g) | N | ε94Mo (± 2σ) | ε95Mo (± 2σ) | Δ95Mo (± 2σ) | Ref. |
|---|---|---|---|---|---|---|---|---|---|---|---|---|
| NWA 2458 | L1 | (70) | 18 | –0.40 ± 0.05 | 0.06 ± 0.04 | –2.44 ± 0.10 | (156) | 1 | 5.30 ± 0.22 | 3.25 ± 0.15 | 9 ± 20 | [1] |
| (L3.2) | L2 | (85) | 18 | –0.18 ± 0.07 | –0.02 ± 0.03 | –1.40 ± 0.09 | (2078) | 6 | 3.18 ± 0.08 | 1.87 ± 0.09 | –3 ± 10 | [1] |
| | L3 | (302) | 14 | –0.07 ± 0.06 | –0.02 ± 0.04 | –0.41 ± 0.07 | (301) | 2 | 1.70 ± 0.22 | 1.00 ± 0.15 | –1 ± 20 | [1] |
| | L4 | (2529) | 25 | –0.22 ± 0.07 | 0.05 ± 0.03 | –0.68 ± 0.06 | (495) | 4 | –2.57 ± 0.28 | –1.82 ± 0.15 | –29 ± 22 | [1] |
| | L5 | (48) | 18 | –0.15 ± 0.08 | –0.04 ± 0.03 | 1.31 ± 0.07 | (737) | 5 | –1.36 ± 0.24 | –0.89 ± 0.14 | –8 ± 20 | [1] |
| | L6 | (60) | 17 | –0.15 ± 0.07 | –0.06 ± 0.05 | –0.40 ± 0.07 | (192) | 2 | –20.53 ± 0.22 | –12.22 ± 0.15 | 2 ± 20 | [1] |
| | *Total / Wtd. Av.* | *(3094)* | | *–0.21* | *0.04* | *–0.67* | *(3958)* | | *0.44* | *0.20* | *–6* | *[1]* |
| WSG 95300 | L1 | (81) | 13 | –0.46 ± 0.06 | 0.01 ± 0.08 | –3.24 ± 0.04 | (98) | 1 | –0.76 ± 0.22 | –0.40 ± 0.15 | 5 ± 20 | [1] |
| (H3.3) | L2 | (101) | 15 | –0.25 ± 0.06 | –0.08 ± 0.03 | –1.54 ± 0.08 | (1198) | 4 | 0.83 ± 0.13 | 0.52 ± 0.04 | 3 ± 9 | [1] |
| | L3 | (132) | 19 | 0.09 ± 0.07 | –0.04 ± 0.03 | 1.66 ± 0.09 | (523) | 4 | –3.30 ± 0.24 | –2.10 ± 0.15 | –13 ± 21 | [1] |
| | L4 | (372) | 18 | –0.21 ± 0.11 | 0.02 ± 0.04 | –0.56 ± 0.07 | (1187) | 6 | –0.10 ± 0.14 | –0.17 ± 0.12 | –11 ± 15 | [1] |
| | L5 | (10) | 13 | –0.03 ± 0.05 | –0.04 ± 0.04 | 0.67 ± 0.06 | (2021) | 6 | 3.13 ± 0.11 | 1.89 ± 0.10 | 2 ± 12 | [1] |
| | L6 | (34) | 13 | –0.15 ± 0.05 | –0.01 ± 0.04 | –0.25 ± 0.08 | (109) | 1 | 6.40 ± 0.22 | 4.32 ± 0.15 | 51 ± 20 | [1] |
| | *Total / Wtd. Av.* | *(731)* | | *–0.19* | *–0.01* | *–0.56* | *(5137)* | | *1.19* | *0.69* | *–1* | *[1]* |
| | *Av. bulk OC [2]* | | | *–0.14 ± 0.02* | *–0.01 ± 0.03* | *–0.62 ± 0.04* | | | *0.66 ± 0.12* | *0.26 ± 0.06* | *–13 ± 9* | |

Titanium isotope ratios are internally normalized to $^{49}Ti/^{47}Ti = 0.749766$ and reported relative to the OL-Ti bracketing standard; Mo isotope ratios are internally normalized to $^{98}Mo/^{96}Mo = 1.453173$ and reported relative to the Alfa Aesar Mo standard. $\Delta^{95}Mo = (\varepsilon^{95}Mo - 0.596 \times \varepsilon^{94}Mo) \times 100$; The $\Delta^{95}Mo$ notation reflects the ppm-deviation of a sample's $^{94}Mo$ and $^{95}Mo$ isotope composition from an *s*-process mixing line passing through the origin in $\varepsilon^{95}Mo$ vs. $\varepsilon^{94}Mo$ space. Given uncertainties of individual samples are Student-t 95% confidence intervals (95% CI) for samples with N>3 [i.e., $(t_{0.95,\ N-1} \times \text{s.d.})/\sqrt{N}$] or reflect the external reproducibility (2 s.d.) obtained from repeated analysis of the terrestrial standards for samples with N≤3 (in case of Mo). For the Ti isotope analyses, solution aliquots (equivalent to ~30 µg Ti) were taken from the previously obtained leachate fractions from Budde et al. (2019). N: number of analyses. References: [1] Budde et al., 2019, [2] Rüfenacht et al., 2023.
Titanium concentrations as determined by quadrupole ICP–MS, which have an uncertainty of ~10%. For the acid leachate fractions the amount of Ti (µg) and Mo (µg) released in each respective leaching step are reported.

**Table 4:** Summary of Ti, Cr, and Mo isotope data for bulk planetary materials.

| | $\varepsilon^{46}$Ti (± 2σ) | $\varepsilon^{50}$Ti (± 2σ) | $\varepsilon^{53}$Cr (± 2σ) | $\varepsilon^{54}$Cr (± 2σ) | $\varepsilon^{94}$Mo (± 2σ) |
|---|---|---|---|---|---|
| ***Carbonaceous chondrites*** | | | | | |
| CI | 0.32 ± 0.05 | 1.91 ± 0.06 | 0.29 ± 0.06 | 1.60 ± 0.08 | 0.79 ± 0.41 |
| CM | 0.51 ± 0.04 | 3.00 ± 0.11 | 0.19 ± 0.04 | 0.96 ± 0.06 | 4.82 ± 0.45 |
| CO | 0.63 ± 0.07 | 3.52 ± 0.44 | 0.10 ± 0.07 | 0.83 ± 0.18 | 1.29 ± 0.34 |
| CV | 0.60 ± 0.04 | 3.39 ± 0.17 | 0.10 ± 0.02 | 0.91 ± 0.05 | 1.27 ± 0.51 |
| CL | 0.49 ± 0.05 | 2.63 ± 0.14 | 0.07 ± 0.05 | 0.70 ± 0.08 | 1.14 ± 0.10 |
| CK | 0.61 ± 0.08 | 3.63 ± 1.05 | 0.07 ± 0.05 | 0.50 ± 0.11 | 1.62 ± 0.14 |
| CR | 0.41 ± 0.06 | 2.22 ± 0.24 | 0.23 ± 0.09 | 1.31 ± 0.07 | 2.86 ± 0.28 |
| CB | 0.33 ± 0.04 | 1.73 ± 0.18 | 0.07 ± 0.26 | 1.28 ± 0.25 | 1.30 ± 0.04 |
| CH | 0.36 ± 0.03 | 2.00 ± 0.08 | 0.16 ± 0.15 | 1.47 ± 0.10 | 1.79 ± 0.10 |
| Tagish Lake | 0.47 ± 0.10 | 2.74 ± 0.06 | 0.37 ± 0.47 | 1.25 ± 0.17 | |
| Tarda | 0.44 ± 0.09 | 2.60 ± 0.09 | 0.23 ± 0.07 | 1.40 ± 0.16 | |
| Bells | | | 0.15 ± 0.01 | 1.23 ± 0.07 | |
| Essebi | 0.48 ± 0.07 | 2.91 ± 0.15 | 0.24 ± 0.05 | 1.44 ± 0.11 | |
| Flensburg | 0.54 ± 0.01 | 3.15 ± 0.47 | 0.20 ± 0.01 | 1.01 ± 0.16 | |
| Ningqiang | | | 0.11 ± 0.03 | 0.75 ± 0.08 | |
| Acfer 094 | 0.43 ± 0.14 | 3.08 ± 0.93 | 0.23 ± 0.06 | 1.39 ± 0.16 | |
| BUC 10933 | | | 0.07 ± 0.04 | 0.86 ± 0.05 | |
| DaG 055 | | 2.84 ± 0.17 | | | |
| DaG 429 | | | 0.02 ± 0.04 | 0.69 ± 0.03 | |
| DaG 430 | 0.48 ± 0.06 | 2.23 ± 0.12 | 0.04 ± 0.02 | 0.75 ± 0.04 | |
| DaG 978 | | | 0.09 ± 0.03 | 0.86 ± 0.08 | |
| EET 83226 | | 4.25 ± 0.15 | | 0.93 ± 0.13 | |
| EET 83355 | | 3.11 ± 0.15 | | 0.76 ± 0.13 | |
| GRO 95566 | | 3.50 ± 0.15 | | 0.92 ± 0.13 | |
| GRO 95577 | 0.32 ± 0.08 | 1.91 ± 0.09 | | | |
| LEW 85332 | | 2.42 ± 0.15 | | 1.23 ± 0.13 | |
| MAC 87300 | | 4.67 ± 0.15 | 0.09 ± 0.02 | 0.73 ± 0.04 | |
| MAC 87301 | | 4.12 ± 0.15 | 0.10 ± 0.03 | 0.82 ± 0.03 | |
| MAC 88107 | | 3.03 ± 0.15 | 0.11 ± 0.03 | 0.90 ± 0.61 | |
| MIL 07513 | | | 0.11 ± 0.03 | 0.93 ± 0.07 | |
| MIL 090292 | | | 0.14 ± 0.02 | 1.26 ± 0.04 | |
| NWA 1152 | | | 0.10 ± 0.02 | 0.72 ± 0.04 | |
| NWA 1839 | | 3.20 ± 0.51 | | 1.03 ± 0.07 | |
| NWA 2994 | | | | 1.31 ± 0.10 | |
| NWA 5958 | 0.48 ± 0.06 | 3.02 ± 0.57 | 0.19 ± 0.08 | 1.17 ± 0.04 | |
| NWA 11024 | 0.47 ± 0.06 | 2.83 ± 0.05 | 0.24 ± 0.06 | 1.02 ± 0.16 | |
| NWA 11086 | 0.44 ± 0.12 | 2.75 ± 0.20 | 0.14 ± 0.16 | 1.03 ± 0.16 | |
| NWA 12416 | 0.45 ± 0.04 | 2.93 ± 0.44 | 0.16 ± 0.04 | 1.29 ± 0.31 | |
| NWA 12957 | 0.44 ± 0.04 | 2.96 ± 0.27 | 0.27 ± 0.04 | 1.05 ± 0.11 | |
| QUE 99038 | | | 0.06 ± 0.01 | 0.92 ± 0.01 | |
| WIS 91600 | 0.47 ± 0.09 | 3.07 ± 0.05 | 0.14 ± 0.08 | 1.33 ± 0.22 | |
| ***Carbonaceous achondrites*** | | | | | |
| Tafassasset | 0.35 ± 0.10 | 2.03 ± 0.06 | 0.21 ± 0.04 | 1.41 ± 0.09 | 1.60 ± 0.16 |
| NWA 2788 | | 2.13 ± 0.51 | 0.00 ± 0.04 | 1.04 ± 0.12 | |
| NWA 2994 | | 2.48 ± 0.96 | | | |
| NWA 2976 | | | | 1.43 ± 0.07 | |
| NWA 3100 | 0.31 ± 0.13 | 1.91 ± 0.31 | | 1.50 ± 0.11 | |
| NWA 3704 | | | 0.39 ± 0.05 | 1.56 ± 0.10 | |
| NWA 6693 | | | 0.31 ± 0.05 | 1.60 ± 0.10 | |
| NWA 6704 | 0.36 ± 0.16 | 2.08 ± 0.01 | | 1.56 ± 0.10 | 1.48 ± 0.12 |
| NWA 6926 | 0.41 ± 0.09 | 1.92 ± 0.13 | 0.26 ± 0.05 | 1.44 ± 0.11 | 1.48 ± 0.12 |
| NWA 7822 | | 2.13 ± 0.51 | | 1.14 ± 0.08 | |
| NWA 8548 | | | | | 1.53 ± 0.10 |
| ***Non-carbonaceous chondrites*** | | | | | |
| EC | –0.05 ± 0.05 | –0.16 ± 0.06 | 0.17 ± 0.03 | 0.02 ± 0.03 | 0.41 ± 0.08 |
| OC | –0.14 ± 0.02 | –0.62 ± 0.04 | 0.22 ± 0.05 | –0.37 ± 0.02 | 0.66 ± 0.12 |
| R | –0.08 ± 0.05 | –0.42 ± 0.07 | 0.23 ± 0.02 | –0.07 ± 0.02 | 0.49 ± 0.32 |
| K | –0.26 ± 0.06 | –1.28 ± 0.08 | 0.56 ± 0.16 | –0.43 ± 0.04 | 0.94 ± 0.12 |
| GRO 95551 | | | 0.01 ± 0.03 | –0.07 ± 0.08 | |
| GRV 020043 | | –1.59 ± 0.24 | | –0.48 ± 0.10 | |
| LAP 04757 | | –0.19 ± 0.15 | | –0.33 ± 0.13 | |
| LAP 04773 | | –0.54 ± 0.15 | | –0.46 ± 0.16 | |
| MIL 15362 | | | 0.18 ± 0.02 | –0.37 ± 0.05 | |
| NWA 5492 | | | | | 0.33 ± 0.16 |
| NWA 5717 | | –0.61 ± 0.07 | 0.22 ± 0.02 | –0.34 ± 0.10 | |
| NWA 13202 | –0.08 ± 0.04 | –0.64 ± 0.05 | 0.28 ± 0.19 | –0.24 ± 0.17 | 0.73 ± 0.16 |
| ***Non-carbonaceous achondrites*** | | | | | |
| Winoniaites | –0.04 ± 0.08 | –0.34 ± 0.20 | 0.34 ± 0.28 | –0.06 ± 0.14 | 0.22 ± 0.10 |
| Aubrites | –0.03 ± 0.07 | –0.06 ± 0.06 | 1.27 ± 0.38 | 0.02 ± 0.07 | 0.53 ± 0.16 |
| HEDs | –0.26 ± 0.03 | –1.24 ± 0.03 | 0.54 ± 0.17 | –0.64 ± 0.08 | |
| Mesosiderites | –0.24 ± 0.04 | –1.24 ± 0.03 | 0.59 ± 0.69 | –0.65 ± 0.11 | 1.03 ± 0.04 |
| Acapulcoites-lodranites | –0.25 ± 0.05 | –1.40 ± 0.06 | 0.27 ± 0.06 | –0.61 ± 0.07 | 0.94 ± 0.30 |
| Angrites | –0.21 ± 0.02 | –1.17 ± 0.07 | 1.04 ± 0.77 | –0.43 ± 0.06 | 0.75 ± 0.11 |
| Brachinites | –0.26 ± 0.09 | –1.33 ± 0.23 | 0.25 ± 0.16 | –0.40 ± 0.15 | 1.15 ± 0.12 |
| Ureilites | –0.36 ± 0.04 | –2.01 ± 0.07 | 0.24 ± 0.07 | –0.86 ± 0.04 | 0.96 ± 0.04 |
| Erg Chech 002 | | –1.08 ± 0.30 | 0.35 ± 0.16 | –0.66 ± 0.23 | |
| GRA 06128 | –0.20 ± 0.06 | –1.58 ± 0.39 | 1.24 ± 0.05 | –0.53 ± 0.28 | 1.03 ± 0.22 |
| GRA 06129 | | –1.55 ± 0.27 | | –0.46 ± 0.13 | |
| NWA 468 | | –1.54 ± 0.42 | | –0.59 ± 0.09 | |
| NWA 1058 | –0.20 ± 0.04 | –0.68 ± 0.10 | 0.50 ± 0.07 | –0.39 ± 0.06 | 1.31 ± 0.11 |
| NWA 2526 | –0.03 ± 0.05 | –0.01 ± 0.13 | 0.17 ± 0.05 | 0.08 ± 0.11 | |
| NWA 5363 | –0.16 ± 0.06 | –1.00 ± 0.14 | 0.23 ± 0.13 | –0.31 ± 0.24 | 0.66 ± 0.22 |
| NWA 5400 | –0.17 ± 0.08 | –1.01 ± 0.05 | 0.13 ± 0.12 | –0.43 ± 0.08 | 0.66 ± 0.22 |
| NWA 6112 | –0.27 ± 0.10 | –1.44 ± 0.09 | 0.24 ± 0.04 | –0.60 ± 0.13 | 1.55 ± 0.22 |
| NWA 7325 | | –1.58 ± 0.33 | | –0.61 ± 0.11 | |
| NWA 8054 | | –1.01 ± 0.38 | | –0.44 ± 0.08 | |
| ***Planets*** | | | | | |
| Earth | –0.05 ± 0.02 | –0.03 ± 0.02 | 0.01 ± 0.02 | 0.03 ± 0.03 | 0.03 ± 0.05 |
| Mars | –0.11 ± 0.03 | –0.41 ± 0.06 | 0.18 ± 0.01 | –0.16 ± 0.03 | 0.50 ± 0.55 |

See data repository: 10.17632/pbdbzbsrvg.1 for a detailed compilation of individual meteorite data and corresponding references.

$\varepsilon^{50}$Ti data: Trinquier et al., 2009; Zhang et al., 2011, 2012; Goodrich et al., 2017; Burkhardt et al., 2017, 2021; Gerber et al., 2017; Bischoff et al., 2017, 2019; Sanborn et al., 2019; Torrano et al., 2019, 2021; Williams et al., 2020, 2021; Metzler et al., 2021; Hellmann et al., 2023; Rüfenacht et al., 2023; Wölfer et al., 2023; this study.

$\varepsilon^{54}$Cr data: Shukolyukov and Lugmair, 2004, 2006; Trinquier et al., 2007, 2008a, b; Qin et al., 2010; Yamakawa et al., 2010; Yamashita et al., 2010; Larsen et al., 2011; Petitat et al., 2011; Schiller et al., 2014; Jenniskens et al., 2014; Göpel et al., 2015; Van Kooten et al., 2016, 2020; Schmitz et al., 2016; Goodrich et al., 2017; Burkhardt et al., 2017, 2021; Mougel et al., 2018; Li et al., 2018; Sanborn et al., 2019; Zhu et al., 2019, 2020, 2021a, b, 2022, 2023; Pedersen et al., 2019; Kadlag et al., 2019, 2021; Bischoff et al., 2019; Kruijer et al., 2020; Williams et al., 2020; Torrano et al., 2021, 2023; Metzler et al., 2021; Anand et al., 2022; Hellmann et al., 2023; Rüfenacht et al., 2023; Krestianinov et al., 2023; Wölfer et al., 2023; Frossard et al., 2024; Kruttasch et al., 2024; Yang et al., 2025; this study.

$\varepsilon^{94}$Mo data: Burkhardt et al., 2011, 2017, 2021; Budde et al., 2016a, 2018, 2019; Render et al., 2017; Worsham et al., 2017; Bermingham et al., 2018; Yokoyama et al., 2019; Hopp et al., 2020; Wölfer et al., 2023; this study.

Supplementary Material

# Origin of nucleosynthetic isotope variability in the NC reservoir: Evidence from Ti, Cr, and Mo isotopes

Elias Wölfer, Christoph Burkhardt, Gerrit Budde, Christian A. Jansen, Jonas Pape, Thorsten Kleine

This file includes:
Supplementary text
Supplementary Tables 1–2
Supplementary Figures 1–5

---

## Disturbed Cr isotope systematics in HaH 193 and NWA 7474

Even though Cr isotope anomalies are mostly nucleosynthetic ($\varepsilon^{54}$Cr) and radiogenic ($\varepsilon^{53}$Cr) in origin, Cr isotope systematics in meteorites can also severely be disturbed by cosmogenic isotope anomalies (Qin et al., 2010; Mougel et al., 2018). For example, Qin et al. (2010) demonstrated that samples with high metal contents (e.g., iron meteorites, metal-rich achondrites, and even some chondrites), low Cr abundances (i.e., a high Fe/Cr ratio), and a long exposure history to galactic cosmic rays (GCR) can significantly be affected in their Cr isotopic composition through spallation reactions on target elements like Fe and Ni that produce Cr isotopes (e.g., $^{56}$Fe(n,α)$^{53}$Cr and $^{61}$Ni(n,2α)$^{53}$Cr). This excess Cr results in a correlated shift of $\varepsilon^{53}$Cr and $\varepsilon^{54}$Cr anomalies towards higher values in a ratio of ~1:4 and can cause extreme Cr isotope anomalies that completely overprint the initial nucleosynthetic and radiogenic isotope variations.

The winonaite sample HaH 193 and the lodranite sample NWA 7474 show strong evidence for anomalous Cr isotope systematics caused by GCR-induced spallation reactions on Fe. For example, HaH 193 exhibits $\varepsilon^{54}$Cr excesses (i.e., $\varepsilon^{54}$Cr = 0.50±0.08; weighted mean calculated using IsoplotR, $n$ = 2) relative to the terrestrial standard that would only be expected for CC meteorites, and moreover, in a plot of $\varepsilon^{50}$Ti vs. $\varepsilon^{54}$Cr, would plot in between the clusters defined by CC and NC meteorites. However, as shown by numerous petrological, chemical, and isotopic studies, winonaites belong to the primitive NC achondrites (e.g., Benedix et al., 1998; Cecchi and Caporali, 2015; Hunt et al., 2017; Worsham et al., 2017; Hopp et al., 2020; Rüfenacht et al., 2023). Of note, the Ti isotopic composition measured for HaH 193 in this study supports an affinity to the NC meteorites as well. In addition to the elevated $\varepsilon^{54}$Cr value, HaH 193 also shows an elevated $\varepsilon^{53}$Cr composition of 0.55±0.03 (weighted mean, $n$ = 2) relative the weighted mean $\varepsilon^{53}$Cr of the other primitive achondrites investigated here ($\varepsilon^{53}$Cr = 0.25±0.04, $n$ = 9). Combined, the elevated $\varepsilon^{53}$Cr and $\varepsilon^{54}$Cr values provide strong evidence for a correlated displacement of Cr isotopes towards higher ε-values as would be expected for GCR irradiated samples. Petrological and bulk chemical data of HaH 193 support this inference. As reported by Hunt et al. (2017), the winonaite sample HaH 193 contains 26.6 vol.% metal, 23.3 wt.% FeO, and 901 µg/g Cr. This results in a Fe/Cr ratio of ~200, which is 2 to 3 times higher than the Fe/Cr typically observed for other winonaites. In addition to the high Fe/Cr, HaH 193 probably also had a long GCR exposure history. Schulz et al. (2012) investigated the effects of cosmic ray exposure (CRE) on non-magmatic iron meteorites and winonaites and found that HaH 193 exhibits the highest thermal neutron fluence measured among all meteorites analyzed in their study. Furthermore, they concluded that surface exposure of HaH 193 and an intensive pre-irradiation of silicates might be possible, given an average exposure age of ~50 Ma for winonaites. As such, there is strong evidence that the measured $\varepsilon^{53}$Cr and $\varepsilon^{54}$Cr values of HaH 193 do not reflect its genuine Cr isotopic composition.

The lodranite sample NWA 7474 also displays $\varepsilon^{53}$Cr and $\varepsilon^{54}$Cr isotope anomalies (i.e., $\varepsilon^{53}$Cr = 0.40±0.09 $\varepsilon^{54}$Cr = –0.23±0.12) that are characterized by $^{53}$Cr and $^{54}$Cr excesses compared to the other acapulcoite-lodranites measured here and to values typically reported for acapulcoite-lodranites in the literature (e.g., Larsen et al., 2011; Goodrich et al., 2017; Li et al., 2018; Rüfenacht et al., 2023). Even though the supposed offset is much smaller than for HaH 193, the correlated excesses in $\varepsilon^{53}$Cr and $\varepsilon^{54}$Cr relative to the mean Cr isotopic composition of the other three acapulcoite-lodranites investigated in this study indicate that the Cr isotope systematics of NWA 7474 may also be affected by cosmogenic effects.

## Correction for cosmogenic effects on Cr isotope systematics of HaH 193 & NWA 7474

Over the past years, different methods have been presented to correct for the effect of cosmogenic isotope variations on Cr (e.g., Shima and Honda, 1966; Trinquier et al., 2007; Qin et al., 2010). However, they all rely on measurements of Cr isotopic compositions on different sample components (e.g., metal and silicate) or on special parameters (e.g., shielding factor, production rate of Cr in a given sample), prerequisites that are not given for the single digested bulk powder of HaH 193. For this reason, a different approach was developed that allows to—at least qualitatively—correct cosmogenic effects on the measured $\varepsilon^{53}$Cr and $\varepsilon^{54}$Cr values of bulk HaH 193. To this end, a two-stage model evolution for the winonaite–IAB iron meteorite parent body that accreted ~1.5 Ma and differentiated ~4.8 Ma after CAI formation (Schulz et al., 2010) is assumed (Fig. S5). Note that for reasons of simplicity, it is supposed that planetary differentiation took place instantaneously as a single-stage event. From the time of differentiation and the inferred $^{55}$Mn/$^{52}$Cr ratio of HaH 193, the present-day $\varepsilon^{53}$Cr isotopic composition of HaH 193 can be calculated and this value is then used to correct the measured $\varepsilon^{53}$Cr value for isotopic variations that are not related to radioactive decay of $^{53}$Mn into $^{53}$Cr. Since cosmogenic effects on $\varepsilon^{53}$Cr and $\varepsilon^{54}$Cr are well-correlated (Qin et al., 2010), the corrected $\varepsilon^{53}$Cr value can ultimately be used to correctly adjust the measured $\varepsilon^{54}$Cr value.

In detail, the applied correction for cosmogenic effects on the Cr isotopic signature of HaH 193 was performed as follows: First of all, it was assumed that the parent body of the winonaites (e.g., the winonaite-IAB parent body) was characterized by chondritic major element ratios and accreted ~1.5 Ma after CAI formation (Schulz et al., 2010, Fig. S5). Based on a study of major and trace element characteristics of several winonaites, Hunt et al. (2017) suggested that the chemical composition of the winonaite parent body was most likely similar to OC, EC, or CM chondrites. Thus, the mean $^{55}$Mn/$^{52}$Cr of 0.71±0.14 reported for these chondrite groups by Trinquier et al. (2008) was taken for calculating the isotopic evolution of the winonaite parent body until its differentiation at ~4.8 Ma after the start of the Solar System (Schulz et al., 2010). The radiogenic $\varepsilon^{53}$Cr isotopic composition of the winonaite-IAB parent body at any time before its differentiation is then given by:

$$\varepsilon^{53}Cr = \left[ \frac{\left( \left(\frac{^{53}Cr}{^{52}Cr}\right)_i + \left(\frac{^{55}Mn}{^{52}Cr}\right)_{spl} \times \left(\frac{^{53}Mn}{^{55}Mn}\right)_i \times \left(1 - e^{(-\lambda \times \Delta t_{CAI})}\right) \right)}{\left(\frac{^{53}Cr}{^{52}Cr}\right)_{std}} - 1 \right] \times 10^4$$

with $\left(\frac{^{53}Cr}{^{52}Cr}\right)_i = \left[\frac{(\varepsilon^{53}Cr)_i}{10^4} + 1\right] \times \left(\frac{^{53}Cr}{^{52}Cr}\right)_{std}$

where $(\varepsilon^{53}Cr)_i$ = –0.23±0.09 is the mean Solar System initial Cr isotopic composition (Trinquier et al., 2008), $(^{53}Cr/^{52}Cr)_{std}$ is the Cr isotopic composition of terrestrial NIST SRM 3112a solution standard, $(^{55}Mn/^{52}Cr)_{spl}$ = 0.71±0.14 is the mean $^{55}$Mn/$^{52}$Cr ratio of OC, EC, and CM chondrites (Trinquier et al., 2008), $(^{53}Mn/^{55}Mn)_i$ = (6.28±0.66) × $10^{-6}$ is the initial Solar System Mn isotopic composition (Trinquier et al., 2008), $\Delta t_{CAI}$ reflects the time after CAI formation in Ma, and $\lambda$ = 0.1873±0.0205 $Ma^{-1}$ is the decay constant of short-lived $^{53}$Mn (Honda and Imamura, 1971).

In the course of the planetary differentiation process, Mn and Cr were fractionated between the hypothetical reservoirs 'metal' or 'core' (i.e., the part that is sampled by the IAB iron meteorites) and 'silicate' or 'mantle' (i.e., the part that is sampled by the winonaites), resulting in sub-chondritic Mn/Cr in what became the metal phase and super-chondritic Mn/Cr in what became the silicate-rich phase. The exact $^{55}$Mn/$^{52}$Cr ratio of the silicate-rich phase used for the subsequent calculation of the radiogenic $\varepsilon^{53}$Cr isotopic evolution of the winonaite HaH 193 source reservoir from the time of differentiation until present-day can be inferred from the Mn and Cr concentrations of HaH 193. Hunt et al. (2017) reported that HaH 193 contains 0.19 wt.% MnO and 901 ppm Cr, corresponding to a stable $^{55}$Mn/$^{52}$Cr isotope ratio of 1.85, when using the stable Cr isotope abundances and atomic weights denoted in Trinquier et al. (2008), and consistent with the super-chondritic Mn/Cr expected form the considerations of the two-stage model evolution. As such, the expected present-day radiogenic $\varepsilon^{53}$Cr isotopic composition of the winonaite sample HaH 193 can be calculated as follows:

$$\varepsilon^{53}Cr = \left[\frac{\left(\left(\frac{^{53}Cr}{^{52}Cr}\right)_{diff} + \left(\frac{^{55}Mn}{^{52}Cr}\right)_{spl} \times \left[\left(\frac{^{53}Mn}{^{55}Mn}\right)_{i} \times e^{\left(-\lambda \times \Delta t_{CAI}\right)}\right] \times \left(1 - e^{\left(-\lambda \times \Delta t_{diff}\right)}\right)\right)}{\left(\frac{^{53}Cr}{^{52}Cr}\right)_{std}} - 1\right] \times 10^{4}$$

where $(^{53}Cr/^{52}Cr)_{diff}$ is the Cr isotopic composition of the winonaite-IAB parent body at the time of planetary differentiation (see above), $(^{55}Mn/^{52}Cr)_{spl}$ = 1.85 is the calculated $^{55}$Mn/$^{52}$Cr ratio of HaH 193, and $\Delta t_{diff}$ is the time after the single-stage differentiation event in Ma. Under the simplistic assumption that planetary differentiation on the winonaite-IAB parent body took place as a single instantaneous event ~4.8 Ma after CAI formation, the calculated present-day Cr isotopic composition of HaH 193 should be $\varepsilon^{53}$Cr = 0.42±0.36, if variations in $\varepsilon^{53}$Cr are solely of radiogenic origin. Note that the reported uncertainty includes all propagated uncertainties induced by the correction procedure, and is relatively large due to the large uncertainty on the initial Solar System $\varepsilon^{53}$Cr isotopic composition amongst others. If only taking the variable $(^{55}Mn/^{52}Cr)_{spl}$ into account [i.e., constant $(\varepsilon^{53}Cr)_i$], the uncertainty on the calculated $\varepsilon^{53}$Cr would be ±0.05. Notwithstanding the above, the calculated absolute value is lower than the measured $\varepsilon^{53}$Cr = 0.55±0.03, thus, providing evidence that the ~13 ppm offset towards higher $\varepsilon^{53}$Cr is probably of cosmogenic origin. Finally, as spallation-induced Cr isotope variations on $\varepsilon^{53}$Cr and $\varepsilon^{54}$Cr correlate in a ratio of 1:4, the measured $\varepsilon^{54}$Cr value of HaH 193 was corrected accordingly. This results in a significant shift (~53 ppm) towards a slightly negative $\varepsilon^{54}$Cr value of –0.03±0.08.

After correction for cosmogenic isotope variations, the Cr isotope composition of HaH 193 is fully consistent with recent analyses of the winonaite sample Tierra Blanca ($\varepsilon^{54}$Cr = –0.14±0.07; Rüfenacht et al., 2023) and with an affinity to the NC meteorites, hence, demonstrating that the approach used here is at least qualitatively appropriate for correcting cosmogenic Cr isotope anomalies. Moreover, the corrected $\varepsilon^{54}$Cr value is close to the terrestrial composition and very similar to the small $\varepsilon^{54}$Cr deficits observed for RC. This is in line with the nucleosynthetic Ti anomalies measured for winonaites and RC in this study, and with Ru and Mo isotopic data from the literature (Worsham et al., 2017; Hopp et al., 2020).

The unusual $\varepsilon^{53}$Cr and $\varepsilon^{54}$Cr excesses of the lodranite sample NWA 7474 were reassessed for CRE-effects by correcting the measured $\varepsilon^{53}$Cr value to the weighted mean value of the other three acapulcoite-lodranite samples (i.e., $\varepsilon^{53}$Cr = 0.32±0.03; $n$ = 3), and analogously correcting the measured $\varepsilon^{54}$Cr value (see above). This results in a corrected $\varepsilon^{54}$Cr = –0.55±0.12, which is in perfect agreement with data reported for acapulcoite-lodranites in the literature (Goodrich et al., 2017; Li et al., 2018; Rüfenacht et al., 2023).

Note that only the corrected Cr isotope data for HaH 193 and NWA 7474 were considered in the discussion part in the main text. Beyond that, it is noteworthy that the induced correction for cosmogenic effects on the Cr isotopic composition of HaH 193 results in a downward shift of ~0.5 $\varepsilon^{54}$Cr, which is about one quarter of the entire spectrum of nucleosynthetic Cr isotope anomalies observed in bulk meteorites and emphasizes the importance of considering such effects for samples with high Fe/Cr. For example, reported Cr isotope data for other metal-rich primitive achondrites (e.g., acapulcoites, lodranites, and ureilites) vary among meteorites within the given group (e.g., ~0.4 $\varepsilon^{54}$Cr-range reported for acapulcoites; Göpel and Birck, 2010; Larsen et al., 2011; Goodrich et al., 2017; Rüfenacht et al., 2023), although their Ti isotopic composition is almost identical (Goodrich et al., 2017; this study). Therefore, Cr isotopic heterogeneity among individual metal-rich achondrites may be related to cosmogenic effects.

## Supplementary Tables

**Table S1:** Summary of the leaching procedure applied to the unequilibrated ordinary chondrites NWA 2458 (L3.2) and WSG 95300 (H3.3).

| Step | Acid volumes | Acid mixture | Temperature | Duration |
|---|---|---|---|---|
| L1 | 25 ml HAc + 25 ml $H_2O$ | 50 ml 8.9 M HAc | 20°C | 1 day |
| L2 | 12.5 ml $HNO_3$ + 25 ml $H_2O$ | 37.5 ml 5.1 M $HNO_3$ | 20°C | 5 days |
| L3 | 15 ml HCl + 17.5 ml $H_2O$ | 32.5 ml 5.0 M HCl | 75°C | 1 day |
| L4 | 15 ml HF + 7.5 ml HCl + 7.5 ml $H_2O$ | 30 ml 14.6 M HF – 2.7 M HCl | 75°C | 1 day |
| L5 | 7.5 ml HF + 7.5 ml HCl | 15 ml 14.6 M HF – 5.5 M HCl | 150°C | 3 days |
| L6a | 14 ml HF + 7 ml $HNO_3$ + 0.4 ml $HClO_4$ | 21.4 ml 19.0 M HF – 5.0 M $HNO_3$ – 2% $HClO_4$ | 180–200°C | 5 days |
| L6b | 14 ml $HNO_3$ + 7 ml HCl | 21 ml 10.3 M $HNO_3$ – 3.7 M HCl | 130–170°C | 3 days |

See Budde et al. (2019) for details.

**Table S2:** Detailed Mo isotope data for the samples of the present study.

| Sample name | Group | Sample type | Weight (mg) | Mo (µg/g) | N | $\varepsilon^{92}$Mo (± 95% CI) | $\varepsilon^{94}$Mo (± 95% CI) | $\varepsilon^{95}$Mo (± 95% CI) | $\varepsilon^{97}$Mo (± 95% CI) | $\varepsilon^{100}$Mo (± 95% CI) | $\Delta^{95}$Mo (± 95% CI) |
|---|---|---|---|---|---|---|---|---|---|---|---|
| LEW 87232 | KC | Whole rock | 526 | 1.41 | 6 | 1.13 ± 0.18 | 0.94 ± 0.12 | 0.37 ± 0.09 | 0.21 ± 0.04 | 0.23 ± 0.06 | –19 ± 11 |
| NWA 13202 (measured) | Ungr. chond. | Magnetic fraction | 168 | 2.39 | 4 | 0.90 ± 0.31 | 0.69 ± 0.16 | 0.39 ± 0.16 | 0.21 ± 0.10 | 0.04 ± 0.22 | –2 ± 19 |
| NWA 13202 (CRE corrected)[a] | Ungr. chond. | Magnetic fraction | – | – | – | 0.97 ± 0.31 | 0.73 ± 0.16 | 0.45 ± 0.16 | 0.22 ± 0.10 | 0.02 ± 0.22 | 1 ± 19 |
| NWA 5492 | GC (grouplet) | Metal separate | 227 | 2.78 | 6 | 0.38 ± 0.17 | 0.33 ± 0.16 | 0.18 ± 0.06 | 0.11 ± 0.04 | 0.05 ± 0.04 | –2 ± 11 |
| GRA 06128 | Ungr. achond. | Whole rock | 549 | 0.10 | 1 | 1.17 ± 0.35 | 1.03 ± 0.22 | 0.60 ± 0.15 | 0.23 ± 0.15 | 0.10 ± 0.22 | –1 ± 20 |
| NWA 13400 | CL | Whole rock | 916 | 1.32 | 9 | 1.68 ± 0.10 | 1.14 ± 0.10 | 0.99 ± 0.06 | 0.52 ± 0.04 | 0.44 ± 0.08 | 31 ± 8 |

The Mo isotope ratios are internally normalized to $^{98}$Mo/$^{96}$Mo = 1.453173 using the exponential law and reported relative to the bracketing Alfa Aesar Mo standard. The reported uncertainties are Student-t 95% confidence intervals (95% CI) for samples with N > 3 or the external reproducibility (2 s.d.) of the bracketing standard for GRA 06128. N: number of analyses.

[a] Molybdenum isotope data is corrected for CRE effects using weighted mean $\varepsilon^{196}$Pt vs. $\varepsilon^{i}$Mo slopes determined for iron meteorite groups (e.g., –0.46±0.07 for $\varepsilon^{92}$Mo, –0.30±0.06 for $\varepsilon^{94}$Mo, –0.37±0.04 for $\varepsilon^{95}$Mo, –0.09±0.08 for $\varepsilon^{97}$Mo, and 0.13±0.06 for $\varepsilon^{100}$Mo; Pape et al., 2024) and are normalized to a pre-exposure $\varepsilon^{196}$Pt of −0.06 (Spitzer et al., 2021).

## Supplementary Figures

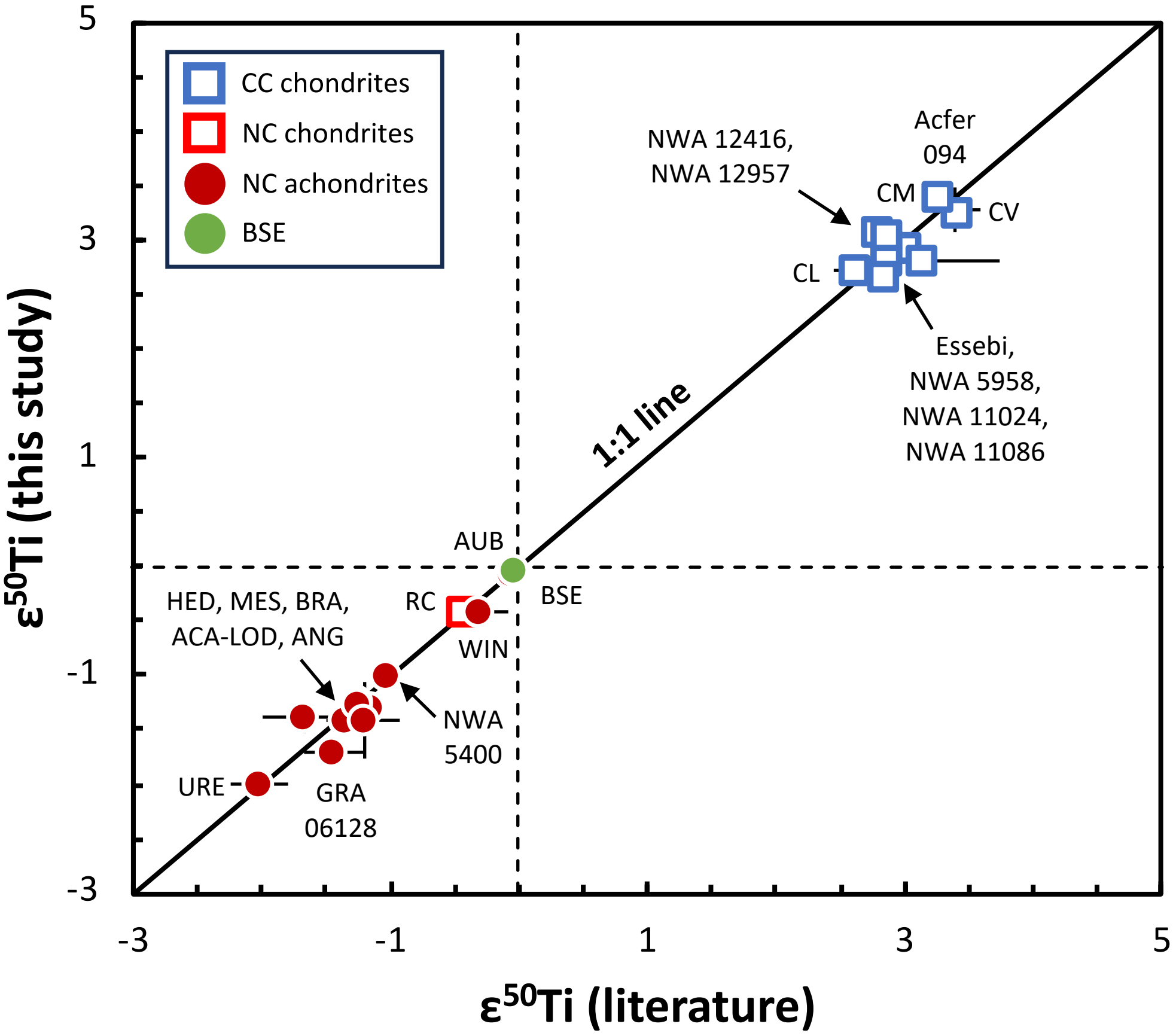

**Fig. S1:** Comparison of the $\varepsilon^{50}$Ti data obtained for various meteorite groups in this study (y-axis) relative to previously reported data (x-axis). Meteorite groups are shown as composite points averaging multiple samples within each group. As such, each data point represents an individual meteorite parent body. References from the literature are given in Table 4 and the data repository. Chondrites are shown by the open squares, achondrites are shown by the filled circles. Abbreviations: Bra – Brachinites, HED – Howardites-Eucrites-Diogenites, Ang – Angrites, Aub – Aubrites, Aca-Lod – Acapulcoite-Lodranites, Win – Winonaites, Ure – Ureilites, Mes – Mesosiderites, RC – Rumuruti chondrites, BSE – Bulk Silicate Earth.

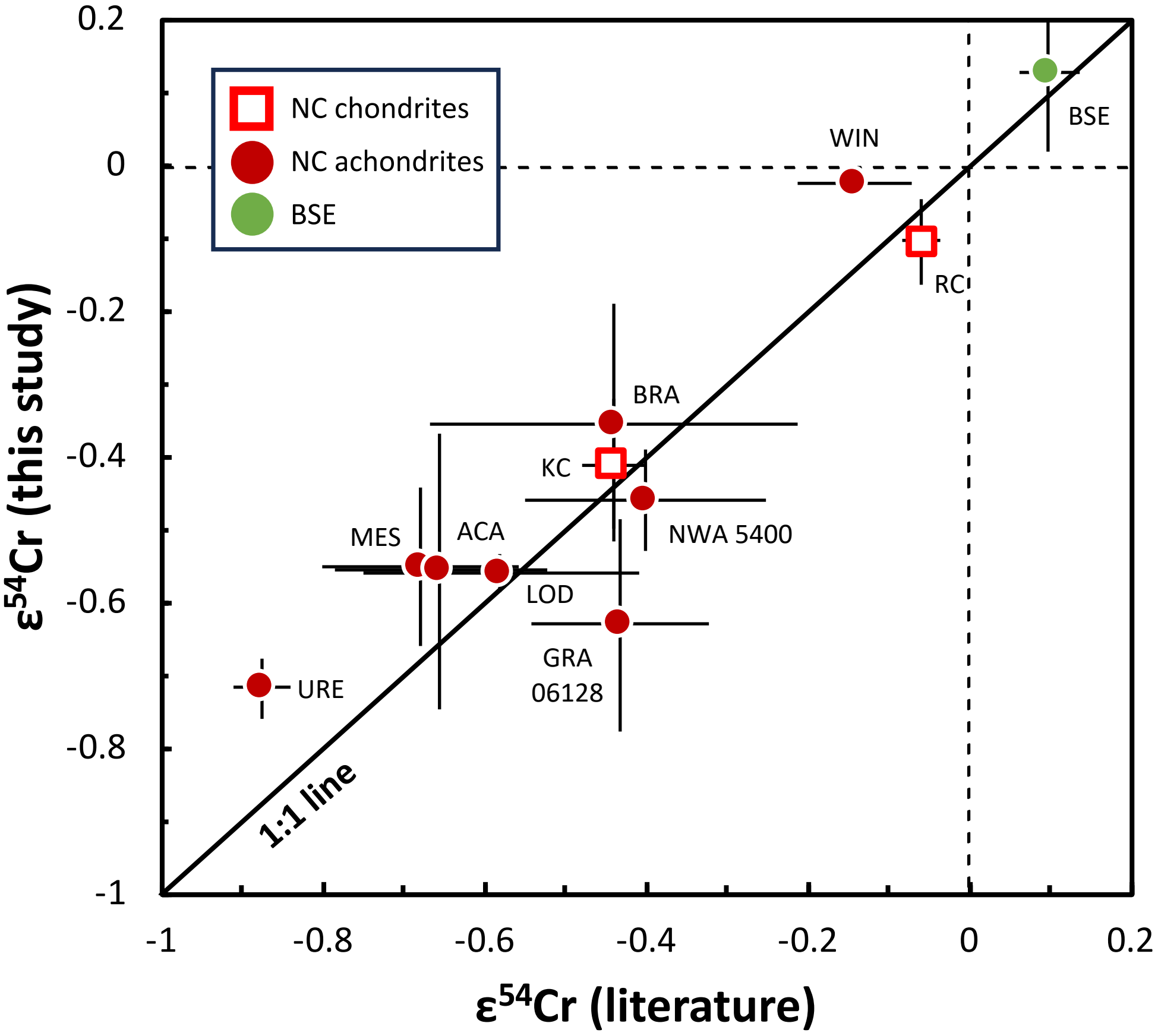

**Fig. S2:** Comparison of the $\varepsilon^{54}Cr$ data obtained for various meteorite groups in this study (y-axis) relative to previously reported data (x-axis). Meteorite groups are shown as composite points averaging multiple samples within each group. As such, each data point represents an individual meteorite parent body. References from the literature are given in Table 4 and the data repository. Chondrites are shown by the open squares, achondrites are shown by the filled circles. Abbreviations: Bra – Brachinites, Aub – Aubrites, Aca-Lod – Acapulcoite-Lodranites, Win – Winonaites, Ure – Ureilites, Mes – Mesosiderites, KC – Kakangari chondrites, RC – Rumuruti chondrites, BSE – Bulk Silicate Earth.

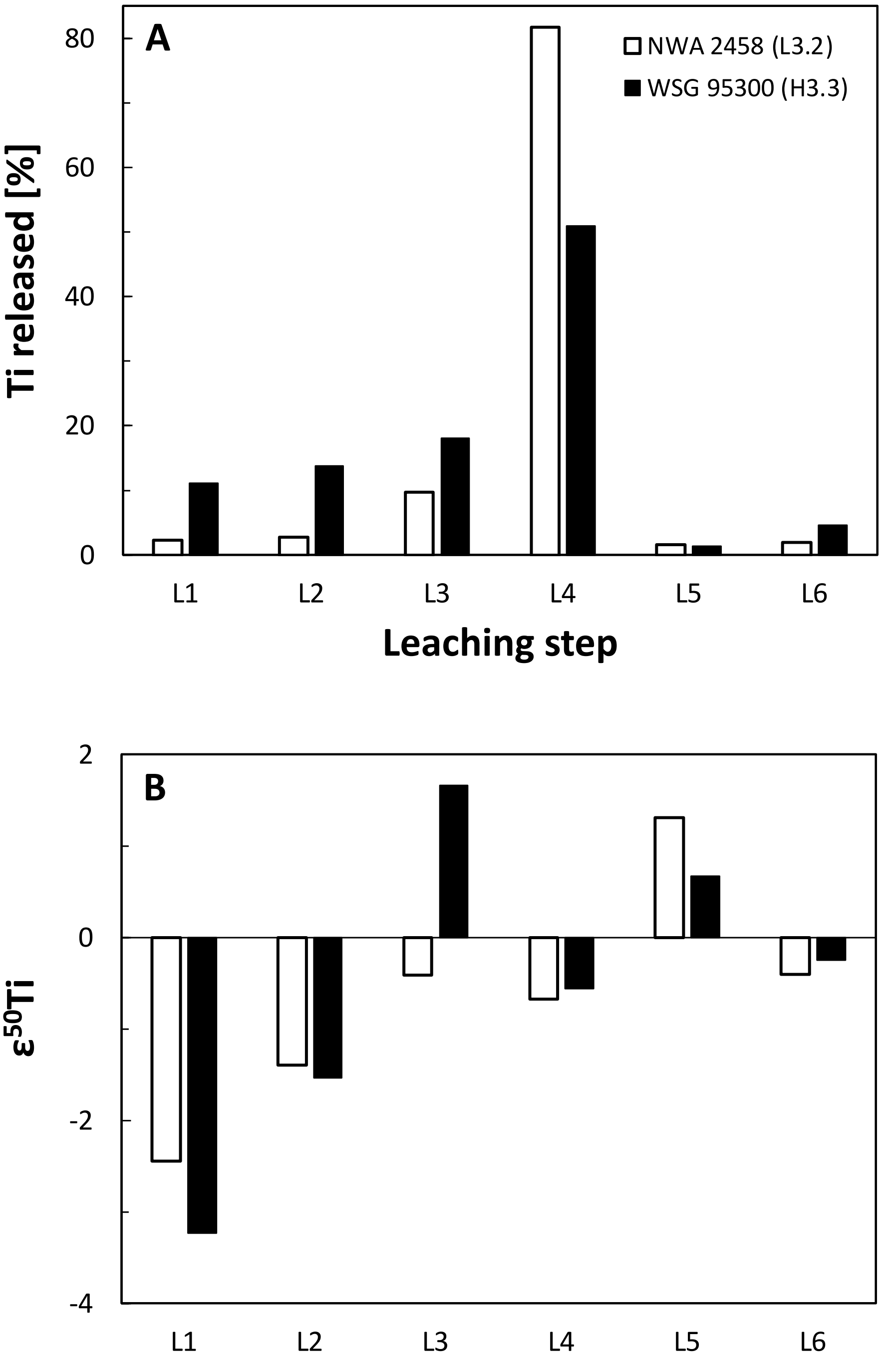

**Fig. S3:** (A) Fraction of Ti released and (B) $\varepsilon^{50}$Ti for the different leaching steps of NWA 2458 (L3.2) and WSG 95300 (H3.3). The two unequilibrated ordinary chondrites show somewhat distinct Ti elemental and isotopic patterns.

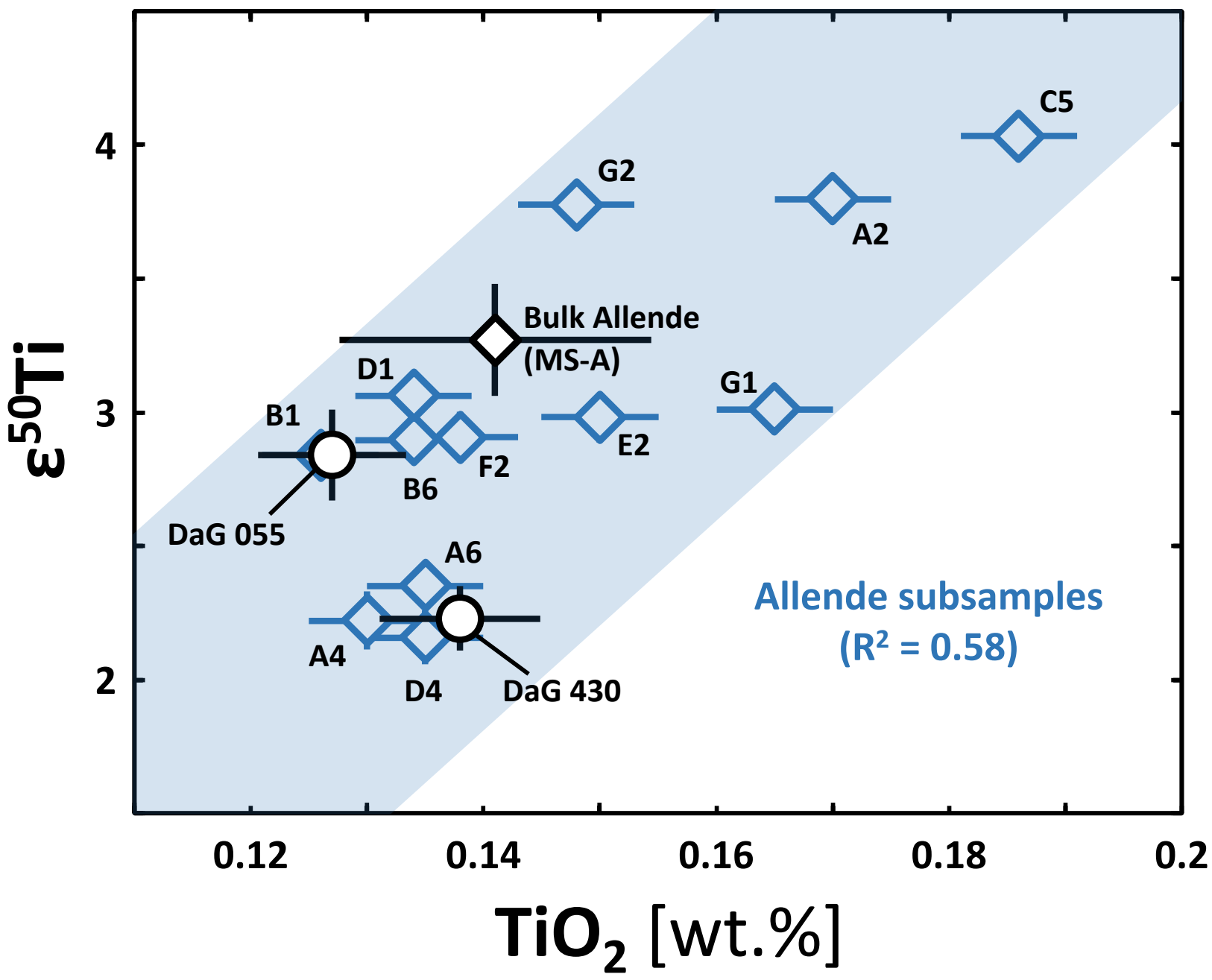

**Fig. S4:** Diagram of $\varepsilon^{50}$Ti vs. $TiO_2$ content of the Allende subsample separates, bulk Allende, and the two ungrouped carbonaceous chondrites DaG 055 and DaG 430. Stated uncertainties on the $\varepsilon^{50}$Ti data reflect 95% CI and are smaller than the symbol sizes in case of the Allende subsamples. The $TiO_2$ data of the Allende subsamples and bulk Allende are from Stracke et al. (2012), those from DaG 055 and DaG 430 are from this study. The blue array reflects a linear regression through the data of the Allende subsamples calculated using IsoplotR ($m$ = 39, $i$ = –2.72). Note that samples with higher $TiO_2$ contents tend to be characterized by larger $^{50}$Ti excesses.

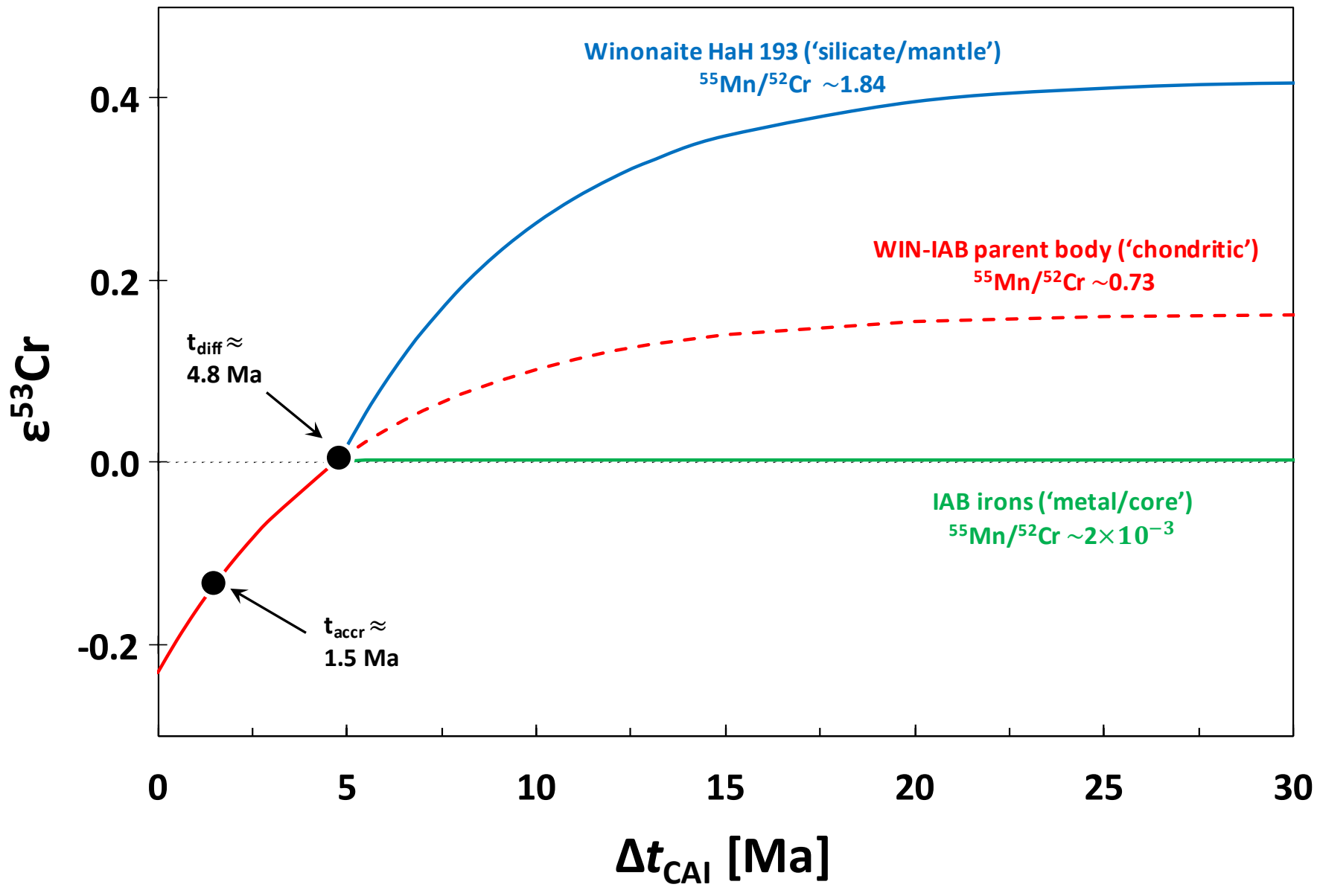

**Fig. S5:** Schematic evolution diagram illustrating the calculation of the present-day $\varepsilon^{53}$Cr isotopic composition of the winonaite sample HaH 193, assuming that metal-silicate separation on the WIN-IAB parent body took place at ~4.8 Ma after CAI formation (Schulz et al., 2010), and was accompanied by significant Mn-Cr fractionation. After differentiation of the WIN-IAB parent body, there is almost no radiogenic $^{53}$Cr ingrowth in the metal phase as sampled by the IAB irons (green line). By contrast, the silicate reservoir as sampled by winonaites evolved with a super-chondritic Mn/Cr towards highly radiogenic $\varepsilon^{53}$Cr isotopic compositions.

## Supplementary materials references

Benedix, G.K., McCoy, T.J., Keil, K., Bogard, D.D., Garrison, D.H., 1998. A petrologic and isotopic study of winonaites: evidence for early partial melting, brecciation, and metamorphism. Geochim. Cosmochim. Acta 62, 2535–2553.

Budde, G., Burkhardt, C., Kleine, T., 2019. Molybdenum isotopic evidence for the late accretion of outer Solar System material to Earth. Nat. Astron. 3, 736–741.

Cecchi, V., Caporali, S., 2015. Petrologic and Minerochemical Trends of Acapulcoites, Winonaites and Lodranites: New Evidence from Image Analysis and EMPA Investigations. Geosciences 5, 222–242.

Goodrich, C.A., Kita, N.T., Yin, Q.-Z., Sanborn, M.E., Williams, C.D., Nakashima, D., Lane, M.D., Boyle, S., 2017. Petrogenesis and provenance of ungrouped achondrite Northwest Africa 7325 from petrology, trace elements, oxygen, chromium and titanium isotopes, and mid-IR spectroscopy. Geochim. Cosmochim. Acta 203, 381–403.

Göpel, C., Birck, J.-L., 2010. Mn/Cr systematics: A tool to discriminate the origin of primitive meteorites? In Goldschmidt Conference Abstracts Goldschmidt Conference 2010. p. 1.

Honda, M., Imamura, M., 1971. Half-Life of $Mn^{53}$. Phys. Rev. C 4, 1182–1188.

Hopp, T., Budde, G., Kleine, T., 2020. Heterogeneous accretion of Earth inferred from Mo-Ru isotope systematics. Earth Planet. Sci. Lett. 534, 116065.

Hunt, A.C., Benedix, G.K., Hammond, S.J., Bland, P.A., Rehkämper, M., Kreissig, K., Strekopytov, S., 2017. A geochemical study of the winonaites: Evidence for limited partial melting and constraints on the precursor composition. Geochim. Cosmochim. Acta 199, 13–30.

Larsen, K.K., Trinquier, A., Paton, C., Schiller, M., Wielandt, D., Ivanova, M.A., Connelly, J.N., Nordlund, Å., Krot, A.N., Bizzarro, M., 2011. Evidence for magnesium isotope heterogeneity in the solar protoplanetary disk. Astrophys. J. 735, L37.

Li, S., Yin, Q.-Z., Bao, H., Sanborn, M.E., Irving, A., Ziegler, K., Agee, C., Marti, K., Miao, B., Li, X., Li, Y., Wang, S., 2018. Evidence for a multilayered internal structure of the chondritic acapulcoite-lodranite parent asteroid. Geochim. Cosmochim. Acta 242, 82–101.

Mougel, B., Moynier, F., Göpel, C., 2018. Chromium isotopic homogeneity between the Moon, the Earth, and enstatite chondrites. Earth Planet. Sci. Lett. 481, 1–8.

Pape, J., Zhang, B., Spitzer, F., Rubin, A.E., Kleine, T., 2024. Isotopic constraints on genetic relationships among group IIIF iron meteorites, Fitzwater Pass, and the Zinder pallasite. Meteorit. Planet. Sci. 59, 778–788.

Qin, L., Alexander, C.M.O., Carlson, R.W., Horan, M.F., Yokoyama, T., 2010. Contributors to chromium isotope variation of meteorites. Geochim. Cosmochim. Acta 74, 1122–1145.

Rüfenacht, M., Morino, P., Lai, Y.-J., Fehr, M.A., Haba, M.K., Schönbächler, M., 2023. Genetic relationships of solar system bodies based on their nucleosynthetic Ti isotope compositions and sub-structures of the solar protoplanetary disk. Geochim. Cosmochim. Acta 355, 110–125.

Schulz, T., Münker, C., Mezger, K., Palme, H., 2010. Hf–W chronometry of primitive achondrites. Geochim. Cosmochim. Acta 74, 1706–1718.

Schulz, T., Upadhyay, D., Münker, C., Mezger, K., 2012. Formation and exposure history of non-magmatic iron meteorites and winonaites: Clues from Sm and W isotopes. Geochim. Cosmochim. Acta 85, 200–212.

Shima, M., Honda, M., 1966. Distribution of spallation produced chromium between alloys in iron meteorites. Earth Planet. Sci. Lett. 1, 65–74.

Spitzer, F., Burkhardt, C., Nimmo, F., Kleine, T., 2021. Nucleosynthetic Pt isotope anomalies and the Hf-W chronology of core formation in inner and outer solar system planetesimals. Earth Planet. Sci. Lett. 576, 117211.

Stracke, A., Palme, H., Gellissen, M., Münker, C., Kleine, T., Birbaum, K., Günther, D., Bourdon, B., Zipfel, J., 2012. Refractory element fractionation in the Allende meteorite: Implications for solar nebula condensation and the chondritic composition of planetary bodies. Geochim. Cosmochim. Acta 85, 114–141.

Trinquier, A., Birck, J., Allegre, C.J., 2007. Widespread $^{54}Cr$ Heterogeneity in the Inner Solar System. Astrophys. J. 655, 1179–1185.

Trinquier, A., Birck, J.-L., Allègre, C.J., Göpel, C., Ulfbeck, D., 2008. $^{53}Mn$–$^{53}Cr$ systematics of the early Solar System revisited. Geochim. Cosmochim. Acta 72, 5146–5163.

Worsham, E.A., Bermingham, K.R., Walker, R.J., 2017. Characterizing cosmochemical materials with genetic affinities to the Earth: Genetic and chronological diversity within the IAB iron meteorite complex. Earth Planet. Sci. Lett. 467, 157–166.